\documentclass{aa} 

\usepackage[breaklinks, colorlinks, citecolor=blue, linkcolor=blue]{hyperref}
\usepackage{graphicx}
\usepackage{placeins}
\usepackage{txfonts}

\begin{document}

   \title{A galactic outflow traced by its extended Mg II emission out to a $\sim30$ kpc radius in the Hubble Ultra Deep Field with MUSE}


   \author{
   Ismael Pessa\inst{\ref{AIP}} \and 
   Lutz Wisotzki\inst{\ref{AIP}} \and 
   Tanya Urrutia \inst{\ref{AIP}} \and 
   John Pharo \inst{\ref{AIP}} \and 
   Ramona Augustin \inst{\ref{AIP}} \and 
   Nicolas F. Bouch{\'e} \inst{\ref{Lyon}} \and 
   Anna Feltre\inst{\ref{INAF}} \and
    Yucheng Guo\inst{\ref{Lyon}} \and
    Daria Kozlova \inst{\ref{AIP}} \and 
   Davor Krajnovic \inst{\ref{AIP}} \and 
   Haruka Kusakabe \inst{\ref{NAOJ}} \and
   Floriane Leclercq\inst{\ref{Texas}} \and
   H\'ector Salas\inst{\ref{AIP}} \and
    Joop Schaye \inst{\ref{Leiden}} \and
    Anne Verhamme \inst{\ref{Geneva}}
   }
   \institute{   
   Leibniz-Institut for Astrophysik Potsdam (AIP), An der Sternwarte 16, 14482 Potsdam, Germany\label{AIP}\\
        \email{ipessa@aip.de}
    \and
    Univ of Lyon1, Ens de Lyon, CNRS, Centre de Recherche Astrophysique de Lyon (CRAL) UMR5574, F-69230 Saint- Genis-Laval, France\label{Lyon}
    \and INAF-Osservatorio Astrofisico di Arcetri, Largo E. Fermi 5, 50125, Firenze, Italy\label{INAF}
    \and National Astronomical Observatory of Japan, 2-21-1 Osawa, Mitaka, Tokyo, 181-8588, Japan\label{NAOJ}
    \and Department of Astronomy, The University of Texas at Austin, 2515 Speedway, Stop C1400, Austin, TX 78712-1205, USA\label{Texas}
    \and Leiden Observatory, Leiden University, PO Box 9513, 2300 RA Leiden, the Netherland \label{Leiden} 
    \and Observatoire de Genève, Université de Genève, Chemin Pegasi 51, 1290 Versoix, Switzerland \label{Geneva}
    }
   \date{Received 29 April 2024; Accepted 28 August 2024}

 
  \abstract
   {We report the discovery of a rare \ion{Mg}{II} $\lambda$$\lambda$ 2796, 2803 doublet emission halo around a star-forming galaxy with $\log (M_\star$/M$_\odot) = 10.3 \pm 0.3$ at $z=0.737$ in deep (9.94 h) VLT/MUSE data from the MUSE-HUDF mosaic. While the central region prominently displays an absorption-dominated \ion{Mg}{II} doublet characterized by discernible P-Cyg features, our examination reveals a remarkably extended \ion{Mg}{II} emission spanning approximately $\sim30$ kpc from the central galaxy. We introduce a simple outflow radiative transfer modeling scheme based on the Sobolev approximation, and we employed a Bayesian Monte Carlo Markov chain fitting to find the best-fitting parameters that match our data. The model reproduces several key features of the observed \ion{Mg}{II} halo and allowed us to constrain the kinematics and geometry of the outflowing gas. Our data are consistent with a biconical wind whose velocity increases with radius, pointing nearly toward the observer, with an opening angle of $59\pm4^{\circ}$. In general, we find that our outflow model performs better in the inner regions of the galactic wind ($\lesssim 10$ kpc $\approx 6$ half-light radii), reaching a velocity of $\sim120$ km s$^{-1}$ at 10 kpc from the central galaxy. However, discrepancies between the data and the model in the outer regions suggest the possible influence of additional mechanisms, such as inflows, satellite interactions, or turbulence, which might significantly shape the circumgalactic medium (CGM) of galaxies at larger impact parameters. This analysis underscores the complexity of galactic outflows and encourages further exploration of the processes governing the dynamics of galactic winds through spatially resolved studies of the CGM.}
   \keywords{galaxies: general --
                galaxies: evolution --
                galaxies: haloes --
                galaxies: structure
               }

\titlerunning{A galactic outflow traced by its extended Mg II emission}
\authorrunning{I. Pessa}
\maketitle

\section{Introduction}

Observing the large-scale gaseous environment of galaxies is a major challenge because of the large projected solid angles covered by galactic haloes and the low densities of the material within \citep{AnglesAlcazar2017, Tumlinson2017, Augustin2019, Corlies2020}. Most constraints come from absorption line studies in privileged lines of sight probed by sufficiently bright background objects, usually quasars \citep[see, e.g.,][]{Bergeron1986,Werk2013}. These have brought up the notion of a generic circumgalactic medium (CGM) where several processes compete to feed galaxies or drive gas out of them \citep[see][for a review on the topic]{Faucher2023}. The spatial distribution of the CGM has been studied mainly through the statistical combination of many independent absorption sight lines. One of the key results of the past years is the clear bimodality of strong low-ionization metal absorption, in particular, the \ion{Mg}{II} $\lambda\lambda$2796, 2803 doublet, with respect to the symmetry axes of the central galaxies, as it indicates a significantly non-isotropic distribution of the circumgalactic gas \citep[e.g.,][]{Bordoloi:2011iu,Bouche:2012ho,Schroetter:2019es}, in agreement with results from cosmological simulations \citep[e.g.,][]{Peroux2020}. For $L^\star$ galaxies, the data are consistent with two main components of the absorbing CGM: a cone-like outflow in the polar direction and an extended disc-like structure close to the equatorial plane \citep{Bouche:2013il,Ho:2017cs,Zabl:2019ija} plausibly related to ongoing gas accretion from the intergalactic medium \citep[although findings from][suggest that cold streams of accreting filaments are not necessarily aligned with the galaxy plane]{Pointon:2019jc}.

However, structural information on individual objects, rather than ensemble averages, is very hard to obtain. With the absorption line technique, this is restricted to rare cases where multiple absorption lines of sight are available for a single galaxy. Even this "tomographic" approach can only provide a very sparse sampling of any particular CGM region. Such measurements have been used to constrain the amount of gas clumping, turbulence, and metallicities on scales of up to a few kiloparsecs \citep[e.g.,][]{Smette:1995wz,Lopez:1999hz,Chen:2014ft,Zahedy:2016cb,Lopez:2018fu,Rubin:2018ja,Peroux:2018is, Afruni2023, Dutta2024} and to test the wind model for \ion{Mg}{II} absorbers \citep{Zabl:2020js}.


Clearly it is desirable to observe the CGM gas directly in emission, especially to perform spectro-mapping of emission lines. Circumgalactic hydrogen Lyman-$\alpha$ (Ly$\alpha$) emission appears to be nearly ubiquitous around low-mass galaxies at high redshifts -- it was first detected statistically in stacked narrowband images \citep{Hayashino:2004ke, Steidel:2011jk, Matsuda:2012fp, Momose:2014fe} and more recently with integral field spectroscopy (IFS) as individual Ly$\alpha$ haloes at $z>3$ \citep{Wisotzki:2016hw, Kusakabe2022, Erb2023}, as well as stacking larger numbers of galaxies \citep{Guo2023b}. These Ly$\alpha$ haloes are typically ten times larger in scale length than the galaxies they surround, and they are detectable out to a substantial fraction of the virial radius \citep{Leclercq:2017cp}. In fact, at the faintest levels, the integrated cross sections of Ly$\alpha$ haloes are comparable to the absorption cross sections of high-column density \ion{H}{I} absorbers, suggesting a close link between cosmic atomic hydrogen found in emission and in absorption \citep{Wisotzki:2018kxa, Kusakabe2022}. The spatially resolved spectroscopy of Ly$\alpha$ haloes has revealed complex line shape variations \citep{Erb:2018dw,Claeyssens:2019cn, Leclercq:2020ht, Erb2023} that are not yet fully understood but that presumably hold key information about the distribution and kinematics of the circumgalactic hydrogen gas. In a recent paper, \citet{Blaizot2023} have been able to reproduce the variety of a large sample of spectroscopically observed Ly$\alpha$ line profiles using the mock spectra of a single galaxy computed for different geometries between the galaxy and the line of sight as well as different evolutionary stages of the galaxy. 

The main challenge here is the resonant nature of the Ly$\alpha$ transition and, at high redshifts, the extreme faintness of circumgalactic emission signals from non-resonant nebular lines, which would be easier to interpret kinematically \citep[although non-resonant lines in the CGM can be studied at low-$z$; see, e.g.,][]{Lokhorst2019}. On the other hand, such lines trace mainly the ionized gas, whereas resonant lines can also probe the neutral gas phase.

At low redshifts, nearly all evidence for the CGM in emission stems from non-resonant "nebular" lines \citep[e.g.,][]{Zhang2016} because of the technical difficulties in measuring spatially extended Ly$\alpha$ at low redshift, with the notable exception of the small sample of Ly$\alpha$ haloes detected by\citet{Hayes:2013jc,Hayes:2014jv} and more recently by \citet{Melinder2023}. This dichotomy of available emission lines -- resonant at $z\ga 2$ and non-resonant at $z\la 1$ -- is an obstacle for joining studies of the CGM in emission obtained at low and high redshifts. In order to bridge these domains, access needs to be gained to both line types for the same galaxies. Apart from the current and planned space-based searches for circumgalactic Ly$\alpha$ emission at intermediate redshifts \citep[e.g.,][]{France:2017bx,Hamden:2019bm,Heap:2019uf}, an interesting approach is to observe emission from the CGM in resonant lines other than Ly$\alpha$ that are accessible to ground-based telescopes at low redshifts.

The \ion{Mg}{II} $\lambda\lambda$2796, 2803 doublet provides possibly the best chances for such measurements. Notably, Mg$^+$ is one of the most abundant metal ions in the universe, especially in the cool-warm neutral and partly ionized gas phase \citep{Tumlinson2017}. 
The resonant nature of the doublet could lead to scattering over large distances and potentially enable extended \ion{Mg}{II} emission similar to Ly$\alpha$ haloes.

The first detection of extended \ion{Mg}{II} emission around a galaxy was reported by \citet{Rubin:2011bm}, who found excess emission over the continuum around a galaxy at $z=0.69$ out to $\sim10$~kpc on both sides. The second and quite similar object, a $z=0.94$ galaxy with a one-sided \ion{Mg}{II} extension of again $\sim 10$~kpc \citep{Martin:2013ho}, was identified as the only such case out of a sample of 145 spectroscopically observed galaxies with coverage of \ion{Mg}{II}. In addition to these two individual detections, \citet{Erb2012} presented statistical evidence that \ion{Mg}{II} emission in a stack of long-slit spectra of star-forming galaxies at $z=1.4$--2 is slightly more extended than the continuum, although the detected excess is very modest. While these studies were hampered by the inevitable limitations of long-slit spectroscopy, inhibiting a complete picture of the distribution of \ion{Mg}{II} emission around the observed galaxies, they suggest that "circumgalactic" \ion{Mg}{II} emission is either a rare phenomenon or very hard to detect (or both). The non-detection of any extended emission around five galaxies in deep narrow-band imaging by \citet{RickardsVaught:2019cx} is consistent with this notion.

In a more recent paper, \citet{Burchett2021} have presented spatially resolved observations from the Keck Cosmic Web Imager (KCWI) of the same galaxy studied by \citet{Rubin:2011bm}. Using integral field spectroscopic data, the authors were able to observe in 3D the circumgalactic \ion{Mg}{II} emission around the central galaxy, and they put some constraints on the physical properties of the CGM. Comparing their data with the radiative transfer models introduced by \citet{Prochaska:2011eqa}, they find it to be consistent with a radially decelerating isotropic wind. This represents the first spatially resolved modeling of a \ion{Mg}{II} halo using integral field spectroscopic data.

In a qualitatively similar study using data from the Multi-Unit Spectroscopic Explorer \citep[MUSE;][]{Bacon:2010jn}, \citet{Zabl2021} reported the first detection of a \ion{Mg}{II} halo in emission, up to distances of 25 kpc from the central galaxy, that is simultaneously probed in absorption by the sight line of a bright background quasi-stellar object (QSO) at an impact parameter of $\sim40$ kpc. The authors find good agreement between the kinematics of the observed emission and absorption, with those produced by a simple toy model of a biconical outflow having a constant velocity, in contrast to the radially decelerating isotropic wind model from \citet{Burchett2021}. 
\\
\citet{Leclercq2022} report the discovery of extended \ion{Mg}{II} emission emerging from the intragroup medium of a group of five star-forming galaxies at $z = 1.31$, with a maximal projected extent of $\sim70$ kpc. They also find evidence suggesting this gas has been ejected through stellar outflows from these galaxies toward the intragroup medium. Moreover, \citet{Leclercq2024} have found evidence of extended \ion{Mg}{II} (and [\ion{O}{II}]) emission in a sample of confirmed Lyman continuum leakers and non-leakers at $z\approx0.35$, aiming to connect the neutral and ionized gas distributions to the Lyman continuum escape fraction.
\\
The work by \citet{Burchett2021}, \citet{Zabl2021}, \citet{Leclercq2022}, and \citet{Leclercq2024} demonstrates how much we can learn about the galactic wind properties through a fully spatially resolved analysis. However, there is still no clear consensus regarding the general properties of galactic winds, such as their velocity, acceleration, precise geometry, outflow mass rate, or loading factor. On the other hand, this kind of analysis requires very deep, high S/N data, and as a consequence, there is only a limited number of examples, which is insufficient to provide a representative view of the CGM in emission. It is clearly necessary to push further in the direction of spatially resolved analyses that preserve the overall structure of the CGM and toward a more robust understanding of the mechanisms that shape it.

In this paper, we present the discovery and analysis of the \ion{Mg}{II} emission halo around a galaxy at $z=0.737$ obtained in observations with the panoramic imaging spectrograph MUSE. We used the high sensitivity and spatial resolution of MUSE to detect \ion{Mg}{II} emission up to distances of more than 20 effective radii from the central galaxy, and then we used a simple outflow modeling scheme to model the inner regions of the halo, allowing us to interpret the observations and put constraints on the properties of galactic outflows. The paper is structured as follows: We summarize our data in Sect.~\ref{sec:obs}, and we give an overview of the properties of our studied galaxy in Sect.~\ref{sec:global}. In Sect.~\ref{sec:mg2}, we emphasize the spatially resolved characteristics of the \ion{Mg}{II} emission and absorption in this galaxy. In Sect.~\ref{sec:model}, we introduce the model used to interpret our observations, and in Sect.~\ref{sec:fitting_method}, we describe the methodology to fit our data. We describe our modeling results in Sect.~\ref{sec:fitting_results}, and in Sect.~\ref{sec:discussion}, we discuss them together with the limitations of our approach and put our results in the context of previous findings. Finally, in Sect.~\ref{sec:summary}, we summarize our results and present some conclusions and perspectives. We used a flat concordance cosmological model with $H_0 = 68$~km~s$^{-1}$ Mpc$^{-1}$ and $\Omega_\mathrm{m} = 0.3$. All quoted transverse distances are in physical units, not co-moving.

\section{Observational data}
\label{sec:obs}

This study is based on observations with the MUSE instrument on the ESO-VLT. MUSE is a panoramic integral-field spectrograph with a field of view (FoV) of $1'\times 1'$ (in Wide Field Mode, used for these observations) at a spatial sampling of $0\farcs2\times 0\farcs2$. The spectral range is 4750~\AA--9350~\AA, with a spectral sampling of 1.25~\AA\ and a wavelength-dependent velocity resolution between $\sim$190 and $\sim$80~km~s$^{-1}$ (full width at half maximum; FWHM) at the blue and red ends of the spectrum, respectively. 

Between 2014 and 2016 we obtained a $3'\times 3'$ MUSE mosaic in the Hubble Ultra Deep Field (HUDF). The effective integration time varies slightly with the position due to the dithering and coadding process; for the region of interest in this paper it is 9.94 hours. The observations and subsequent data processing of the MUSE Mosaic in the HUDF are described in detail in \citet{Bacon:2017hn} and \citet{Bacon2023}. The data are seeing-limited, with a resolution (FWHM of the best-fitting \citet{Moffat:1969ts} function) of 0\farcs63 at 7000~\AA\ (0\farcs71 at 4850~\AA, the wavelength of the \ion{Mg}{II} line in the main galaxy of this study). A first catalog with identifications and redshift measurements of 1439 objects in the MUSE-HUDF mosaic was published in \citet{Inami:2017bm}. The data has then been reprocessed with improved flat fielding and sky subtraction, followed by a second pass through the entire dataset using several new or updated tools; the outcome of this reprocessing has been published in \citet{Bacon2023} as part of the second data release of the MUSE-HUDF\footnote{Publicly available for download in the AMUSED website \url{https://amused.univ-lyon1.fr/}.}. The natural coordinate system of the MUSE data cube is tilted by $-44$\degr\ with respect to the north-south direction.

In addition to the MUSE observations, we employed the extensive multiwavelength data available for the HUDF via public archives, in particular the deep optical and near-infrared images obtained with the \textit{Hubble} Space Telescope \citep[HST;][]{Beckwith:2006hp,Illingworth:2013dk,Oesch:2018eo}, mid-infrared photometry by the \textit{Spitzer} Space Telescope \citep{Labbe:2015dl}, and X-ray observations of the \textit{Chandra} satellite \citep{Luo:2017be}.

\begin{figure}
\includegraphics[width=\columnwidth]{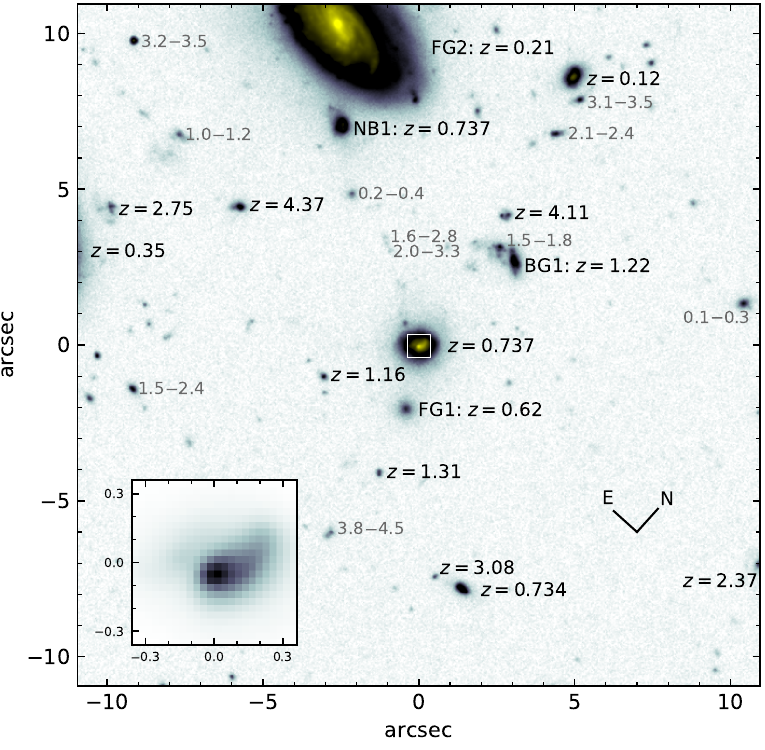}
\caption[]{Image from HST/ACS of the galaxy MUSE-UDF~\#884 (in the field center) and its environment in the F775W filter. The central parts of the brightest galaxies are shown at higher cut levels in yellow. The inset in the lower left provides an enlarged view of the image center, indicated by the white square. Spectroscopic redshifts measured with MUSE are labeled in black, always referring to the object left of the label. The gray labels give photometric redshift ranges taken from \citet{Rafelski:2015iu}, estimated with the BPZ code. Galaxies mentioned in the text are marked as `NB' (physical neighbor), `FG' (foreground) and `BG' (background). The image orientation follows that of the MUSE data cube and is given by the compass marker on the lower right.
}
\label{fig:hst}
\end{figure}

\section{Global properties of the galaxy}
\label{sec:global}

\subsection{Discovery and environment}
\label{sec:discenv}

As part of the above-mentioned reprocessing of the MUSE-HUDF Mosaic data, we searched again for emission line peaks in the data cube, using the LSDCat software \citep[Line Source Detection and Cataloguing, ][]{Herenz:2017er}. This blind search revealed two significant peaks, above our signal-to-noise (S/N) detection threshold of 5.0, which turned out to be a partial ring of \ion{Mg}{II} $\lambda$2796 emission (with its corresponding \ion{Mg}{II} $\lambda$2803 counterpart) surrounding a previously known galaxy at $z=0.737$. At the redshift of this galaxy, 1\arcsec\ corresponds to $7.5$ kpc.


The central galaxy itself has cross-identifications%
\footnote{e.g., ID \#23061 in UVUDF \citep{Rafelski:2015iu}, ID \#13337 in CANDELS \citep{Guo:2013ig}, ID \#25592 in 3D-HST \citep{Skelton:2014do}}
 in many surveys of the HUDF/CDFS region. It is listed as ID \#884 in the catalog of the MUSE-HUDF mosaic by \citet{Inami:2017bm}, and for brevity, we will refer to it by only this number in the following. It is a luminous blue galaxy with $m_{\mathrm{AB}} = 21.7$ in the ACS/F775W band, implying $L\approx L^\star$ at $z=0.737$. Figure~\ref{fig:hst} shows a deep HST image of the galaxy and its immediate surroundings within $\pm 11\arcsec$ ($\pm$80~kpc). Several of the galaxies close to the line of sight have redshifts measured with MUSE and are included in the \citet{Inami:2017bm} sample; these are labeled in the figure. We also provide a few photometric redshifts \citep[taken from][]{Rafelski:2015iu} where this may be of interest. Galaxies unlabelled in Fig.~\ref{fig:hst} have no reliable redshift information or are most likely unrelated to \#884. The physically closest galaxy -- marked as `NB1' -- is MUSE-HUDF \#6516 ($m_{\mathrm{AB}} = 23.8$) at 53~kpc projected distance and with only $-85$~km~s$^{-1}$ redshift difference. Other than this, there are no surrounding objects that are plausibly located nearby in real space; the galaxy at $z=0.734$ near the bottom of the field has a redshift separation of $-1000$~km~s$^{-1}$ and is thus probably already in the foreground. On transverse scales of up to $\sim 1$~Mpc, the full MUSE-HUDF Mosaic reveals 23 (11) galaxies within $\pm$ 1000 (300) km~s$^{-1}$, most of them faint emission line objects scattered all over the field. There is no evidence to support \#884 being located in a cluster or rich group environment. 
 
Other projected neighbors to \#884 are marked in Fig.~\ref{fig:hst} as `BG' (background) and `FG' (foreground). Since these galaxies are faint, we used the TDOSE software package \citep[Three-Dimensional Optimal Spectral Extraction,][]{Schmidt:2019he} to perform an optimal spectra extraction. Each of the galaxies was modelled in the HST data as a superposition of multiple components, which after blurring to the MUSE point spread function (PSF) were then used to define the spatial profiles of the galaxies used in the extraction. We briefly comment on three of these projected neighbors:

\begin{itemize}\setlength{\itemsep}{0ex}

\item BG1 (= MUSE-HUDF \#6943 in \citealt{Inami:2017bm}) is a spectro\-scopically confirmed system at $z=1.221$, at 4\arcsec\ distance to \#884, and with $m_{\mathrm{AB, F775W}} = 24.5$ it is sufficiently bright to produce a well-detected continuum in the MUSE spectrum. The continuum-subtracted data cube shows two significantly negative peaks -- indicative of absorption in the subtracted continuum -- at the location of BG1 and at the wavelengths of the \ion{Mg}{II} emission around \#884. We analyze and discuss this interesting sight line in Sect.~\ref{sec:bg1abs}. 

\item The galaxy marked as FG1 ($m_{\mathrm{AB, F775W}} = 25.3$) has a projected distance of only 2\arcsec\ to \#884, but no redshift in \citet{Inami:2017bm}. The photometric redshift estimate in \citet{Rafelski:2015iu} puts it in the foreground, $0.38 < z_{\mathrm{BPZ}} < 0.53$. The extracted MUSE spectrum does not show any emission features, apart from [\ion{O}{II}] spilling over from \#884, and it is too noisy to allow for a confident identification of individual absorption lines. It, however, agrees well in shape with a synthetic galaxy spectrum from the \citet{Jimenez2004} spectral library for an age of $\sim 2$~Gyr at a redshift of $z = 0.621\pm 0.002$. This redshift is outside the range estimated for the photometric redshift. However, assessing the reliability of the \citet{Rafelski:2015iu} catalog is outside the scope of this paper. While our redshift measurement is not as secure as those based on multiple individual spectral lines, we consider it sufficiently robust to conclude that this galaxy is indeed located in the foreground of \#884 and thus cannot provide another absorption line of sight. 

\item The very bright foreground galaxy FG2 at $z=0.21$ (MUSE-HUDF \#855, $m_{\mathrm{AB, F775W}} = 18.6$) is of concern for this investigation only in that it heavily contaminates the spectrum and immediate surroundings of NB1.

\end{itemize}

All other background galaxies shown in Fig.~\ref{fig:hst} are too faint to produce any detectable continuum signal in our MUSE data cube.

\begin{figure*}
\includegraphics[width=\textwidth]{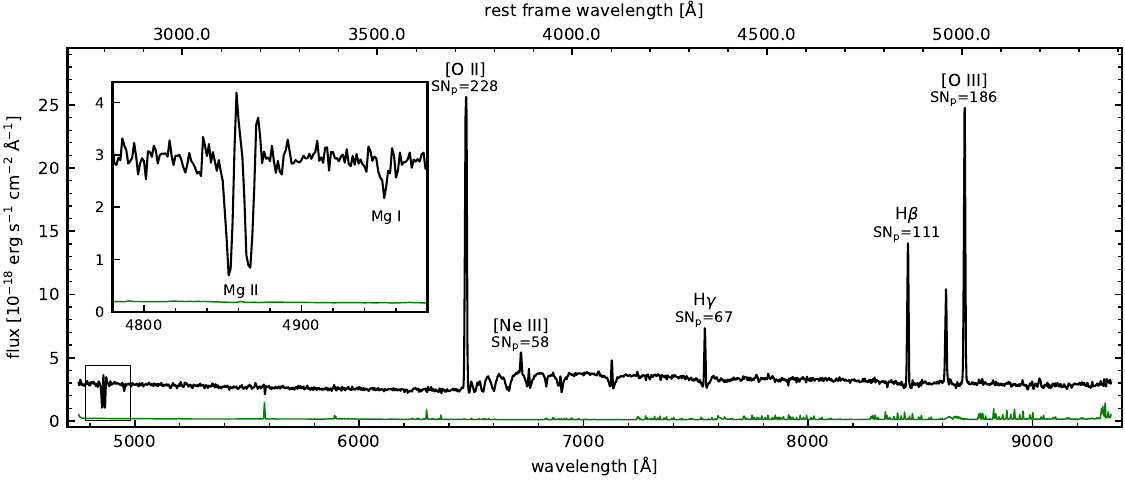}
\caption[]{Integrated MUSE spectrum of the galaxy \#884 within an aperture of 1\arcsec\ radius, slightly smoothed for display purposes. The \ion{Mg}{II} $\lambda\lambda$2796, 2803 doublet and the neighboring \ion{Mg}{I} $\lambda$2852 line are indicated by the box; an unsmoothed zoom into this spectral region is shown in the inset panel. The green line at the bottom traces the $1\sigma$ uncertainty spectrum and indicates the locations of bright sky emission. Other prominent emission lines in the spectrum are labeled, with their corresponding peak signal-to-noise ratio SN$_{\rm p}$.}
\label{fig:spec}
\end{figure*}

\begin{table}
\caption[]{Global properties of the galaxy MUSE-HUDF \#884.}
\begin{tabular}{@{}ll}
\hline\noalign{\smallskip}
Coordinates (J2000.0) & 03:32:44.20, \ $-$27:47:33.5 \\
Systemic redshift & $z = 0.73722 \pm 0.00003$ \\
Absolute magnitude  & $M_{\mathrm{AB}} = -20.7$ \\
Stellar mass & $\log (M_\star$/M$_\odot) = 10.3 \pm 0.3$ \\
Star formation rate & $\mathrm{SFR} \simeq 10\pm7\:\mathrm{M}_\odot\:\mbox{yr}^{-1}$ \\
Half-light radius (F125W) $^\mathrm{a}$ & $r_\mathrm{e} = (1.45 \pm 0.01)\:\mathrm{kpc}$ \\
Sersic index (F125W) $^\mathrm{a}$ & $n = 1.82 \pm 0.02$ \\
\noalign{\smallskip}\hline\noalign{\smallskip}
\multicolumn{2}{l}{$^\mathrm{a}$ from \citet{VanderWel:2012eu} }
\end{tabular}
\label{tab:global}
\end{table}

\subsection{Integrated spectrum}
\label{sec:spec}

In Fig.~\ref{fig:spec}, we present a spectrum of the galaxy extracted by summing the data cube over a circular aperture of 1\arcsec\ radius. The spectrum displays a blue stellar continuum with prominent Balmer absorption and strong nebular emission of [\ion{O}{II}], [\ion{Ne}{III}], the Balmer series, [\ion{O}{III}], and other well-known lines. The spectrum shows \ion{Mg}{II} to be mainly in absorption, but with significant redshifted P-Cygni emission humps as revealed by the inset panel in Fig.~\ref{fig:spec}. Furthermore, \ion{Mg}{I} $\lambda$2852 is detected in absorption.

We fitted the stellar continuum with the pPXF code \citep{Cappellari:2004gm} with the same setup as in \citet{Guerou:2017ik}, using the Indo-US stellar templates \citep{Valdes:2004ba} convolved with the MUSE line spread function (LSF). We obtained a systemic redshift of $z=0.73722$, consistent between stellar absorption and nebular [\ion{O}{II}] emission to within a few kilometers per second. Further derived properties of the galaxy are listed in Table~\ref{tab:global}. The stellar mass and star formation rate (SFR) values have been published by \citet{Bacon2023} as part of the second data release of the MUSE-HUDF. They have been estimated by fitting the broad-band spectral energy distribution with the Prospector code \citep{Prospector}, employing all available bands in the \citet{Rafelski:2015iu} photometric catalog, using the FSPS models \citep{Conroy2009} with the MILES \citep{Vazdekis2015, Vazdekis:2016ho} spectral library and MIST isochrones \citep{Choi2016} to generate synthetic spectra to match the observations. These models include contributions from nebular continuum and emission lines as described by \citet{Byler2017}. An exhaustive description of the SED modeling is available in \citet{Bacon2023}. With these values of stellar mass and star formation rate, the galaxy is located somewhat above the ridge line of the so-called "star formation main sequence" at $z\simeq 0.7$ \citep[e.g.][]{Noeske:2007kq, Lee:2015br, Boogaard:2018hc}. The star formation rate estimate is also consistent with the value obtained from the H$\beta$ spectral line measurement (from the integrated spectrum).


The emission line spectrum of \#884 is that of a normal star-forming galaxy of moderate excitation level. The line widths also reveal no evidence for an AGN, nor do we find any AGN-characteristic emission line such as [\ion{Ne}{V}] $\lambda$3426 in the spectrum. This is in agreement with the marginal detection ($5\sigma$, in the soft band only) as an X-ray source in the 7~Ms exposure of the \textit{Chandra} observatory \citep{Luo:2017be}. The X-ray luminosity of $2\times 10^{41}$~erg~s$^{-1}$ is an order of magnitude below the limit usually adopted as the criterion for AGN, in line with the soft X-ray spectrum. Given the observed star formation rate, it is also consistent with the expected X-ray emission arising from stellar processes alone \citep{Mineo:2014cc}. While these diagnostics do not exclude the possibility of a weak AGN in the center of the galaxy, all indicators suggest that any such source would have to be energetically subdominant.


\begin{figure}
\includegraphics[width=\columnwidth]{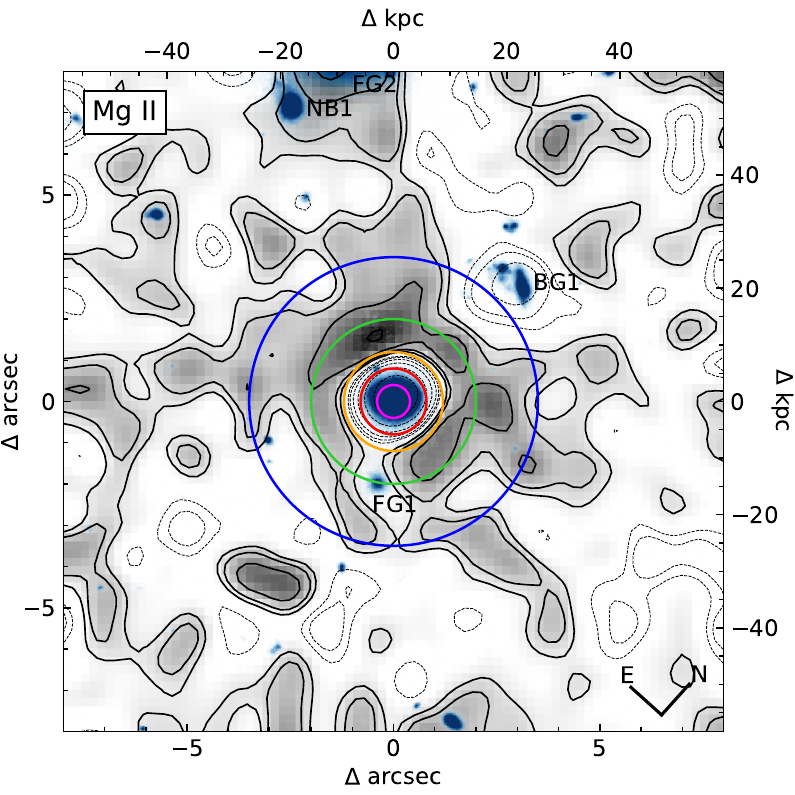}
\caption[]{Pseudo-narrowband \ion{Mg}{II} image extracted from the MUSE data cube by summing over the spectral range of the doublet and filtering with a Gaussian of 1\arcsec\ FWHM. Underlying in blue scale is the HST F775W image, with other galaxies labeled as in Fig.~\ref{fig:hst}. The black solid contours are spaced logarithmically at 0.25~dex interval, starting at $\log_{10}(s)$ of $-18.5$, where $s$ is surface brightness in erg s$^{-1}$ cm$^{-2}$ arcsec$^{-2}$. The lowest contour corresponds to a S/N of $\sim$1.5 in the filtered image. The dotted contours indicate negative measured values of $s$. The central hole in the \ion{Mg}{II} surface brightness distribution is a result of the continuum subtraction, since \#884 is a net absorber in its central spectrum. Similarly, the apparently negative region at the background galaxy BG1 is caused by the strong \ion{Mg}{II} absorption in the continuum of BG1. The colored circumferences indicate the radii of the annular apertures used to extract radially resolved spectra in Sect.~\ref{sec:radial}. 
}
\label{fig:mg2nb1}
\end{figure}

\section{\ion{Mg}{II} absorption and emission}
\label{sec:mg2}

\subsection{\ion{Mg}{II} narrowband image}
\label{sec:nb}

Figure~\ref{fig:mg2nb1} reveals the extended nature of the \ion{Mg}{II} emission. We constructed this pseudo-narrowband image by summing the continuum-subtracted data cube over the entire wavelength range of the \ion{Mg}{II} doublet, 2790--2810~\AA\ in the rest frame of the galaxy. The continuum cube was constructed using a running median across the spectral dimension of the original cube, within a window of 151 wavelength channels. There are no other strong spectral features in the immediate vicinity of \ion{Mg}{II} that could lead to artifacts in the inferred continuum.


The \ion{Mg}{II} emission can be clearly traced out to distances of 30--40~kpc, more than $20\:r_\mathrm{e}$ from the galaxy. The emission is somewhat elongated in shape, but surrounds the galaxy on all sides. We emphasize that the ring-like appearance of the emission around the object is a consequence of the continuum subtraction procedure: Since \#884 is a net absorber in its central spectrum, these pixels become negative after subtracting the continuum. Nevertheless, the \ion{Mg}{II} emission is also present in the central region. An analysis of different emission line ratios that will be presented in a future paper (Wisotzki et al. in prep.), showing that the \ion{Mg}{II} emission halo is consistent with being produced by the resonant scattering of continuum photons emitted by the central source with the ions present in the CGM of \#884.


Similarly to the central depression, the \ion{Mg}{II} emission in Fig.~\ref{fig:mg2nb1} also seems to be avoiding the galaxy BG1, and possibly also FG1. While for the former the explanation is again strong \ion{Mg}{II} absorption in the underlying continuum of BG1, this is not possible for FG1 as the object is located in the foreground. However, any intrinsic stellar absorption feature in the FG1 spectrum that happens to be located within the \ion{Mg}{II} narrowband range will have a similar effect. Since BG1 is sufficiently far away from \#884 in terms of projected distance ($\sim30$ kpc), we keep  BG1 and its immediate surroundings outside the region where we model the \ion{Mg}{II} emission (see Sect.~\ref{sec:fitting_results}).

At the upper edge of the image we see the bright foreground galaxy FG2 and the physical neighbor NB1. Most of the apparent signal in these spaxels is due to continuum-subtraction residuals from FG2 that are not restricted to the \ion{Mg}{II} wavelength range. Similarly to BG1, this galaxy is also beyond the modeled region, to avoid our results being influenced by contaminant emission.

On the other hand, we find that NB1 is not surrounded by any large-scale extended \ion{Mg}{II} emission (its apparent halo in Fig.~\ref{fig:mg2nb1} is an artifact generated by the residual of the continuum subtraction of FG2). The spectrum of NB1 shows only weak \ion{Mg}{II} emission, with a 2-to-1 ratio between the two \ion{Mg}{II} emission lines, suggesting a nebular origin. However, it is certainly possible that \#884 and NB1 are in gravitational interaction, and that this interaction could influence the morphology of the \ion{Mg}{II} halo through tidal forces, in the sense that it shows its largest extent along the direction toward NB1 \citep[see, e.g.,][]{Leclercq2022}.
\\
\begin{figure*}
\includegraphics[width=\textwidth]{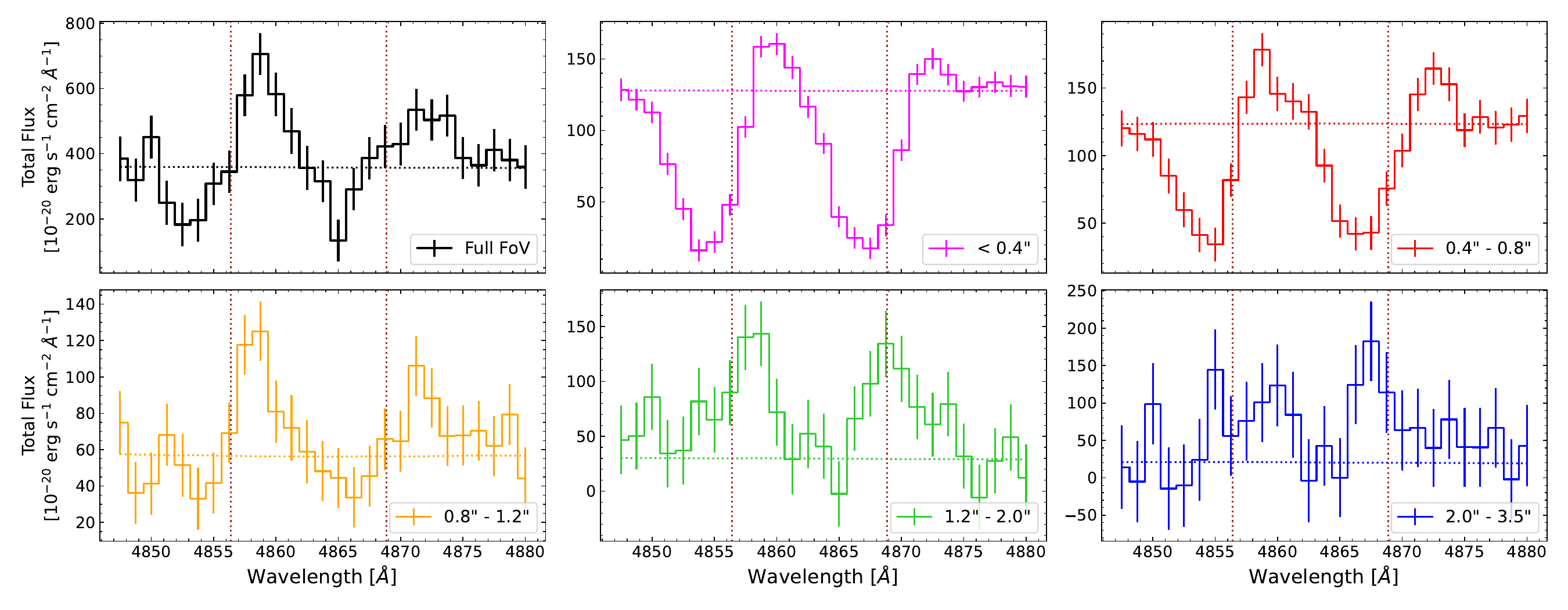}
\caption[]{Spatially resolved \ion{Mg}{II} spectra of galaxy \#884, extracted as the sum of spaxels in circular concentric ring-like apertures (circular aperture in the case of the central region). Each panel corresponds to a different annulus, as indicated in the label, except for the top-left panel, which corresponds to the spectrum of the full modeled FoV. In the frame of reference of the central galaxy, the apertures represent physical sizes of $<3$ kpc, $3-6$ kpc, $6-9$ kpc, $9-15$ kpc, and $15-26$ kpc. The error bars represent the $\pm 1\sigma$ uncertainty of the spectra. The nearly horizontal dotted lines indicate the approximate continuum level. The vertical thin dashed lines mark the observed wavelengths of the \ion{Mg}{II} doublet at the systemic redshift of the galaxy.}
\label{fig:aperspec1}
\end{figure*}

\subsection{Radially resolved spectra}
\label{sec:radial}

In Fig.~\ref{fig:aperspec1}, we show spectra of the \ion{Mg}{II} doublet, extracted in five disjoint concentric circular annular apertures, defined by the radii $<0\farcs4$, $0\farcs4 - 0\farcs8$, $0\farcs8 - 1\farcs2$, and $1\farcs2 - 2\farcs0$, or $<3$ kpc, $3 - 6$ kpc, $6 - 9$ kpc,  $9 - 15$ kpc, and $15 - 26$ kpc, in the frame of reference of the central galaxy, as well as for the full modeled FoV. These apertures are shown in Fig.~\ref{fig:mg2nb1}, following the same color scheme as in Fig.~\ref{fig:aperspec1}. The central spectrum (magenta) is dominated by absorption with only very weak P-Cygni emission humps. The emission then gets progressively stronger compared to the absorption when moving outward. At $1\farcs2 < r < 2\farcs0$ and beyond the absorption is essentially undetected.

The P-Cygni line profile of the \ion{Mg}{II} doublet in the central spectrum of \#884 is quite similar to what has been reported for other star-forming galaxies \citep[e.g.,][]{Weiner:2009cf,Rubin:2011bm,Erb2012,Feltre:2018in}. Such line shapes are often interpreted as signatures of resonant scattering in an optically thick outflow, with strong absorption in the approaching (blueshifted) regions of the flow and a redshifted emission component due to backscattering in the receding material \citep[e.g.,][]{Castor1970, Castor1979, Scuderi1992}. 

The two absorption components of the doublet have approximately equal strength, implying heavy saturation with optical depth $\tau \gg 1$. The absorption appears to get weaker for wider apertures, but this is partially caused by adding \ion{Mg}{II} emission from the outer regions which fills the absorption lines in the continuum.

\section{Model}
\label{sec:model}
Having established the extended nature of the \ion{Mg}{II} emission, we now employ an outflow modeling scheme to interpret our data further. For simplicity, we opt to use the Sobolev approximation \citep{Ambartsumian1958, Sobolev1960, Grinin2001} to describe the interactions between scattered continuum photons and ions in the outflowing material, instead of employing full radiative transfer calculations. Despite this simplification, our model is able to reproduce many key features observed in our data.

\begin{figure*}
\centering
\includegraphics[trim={4cm 0 4cm 0}, width=0.5\textwidth]{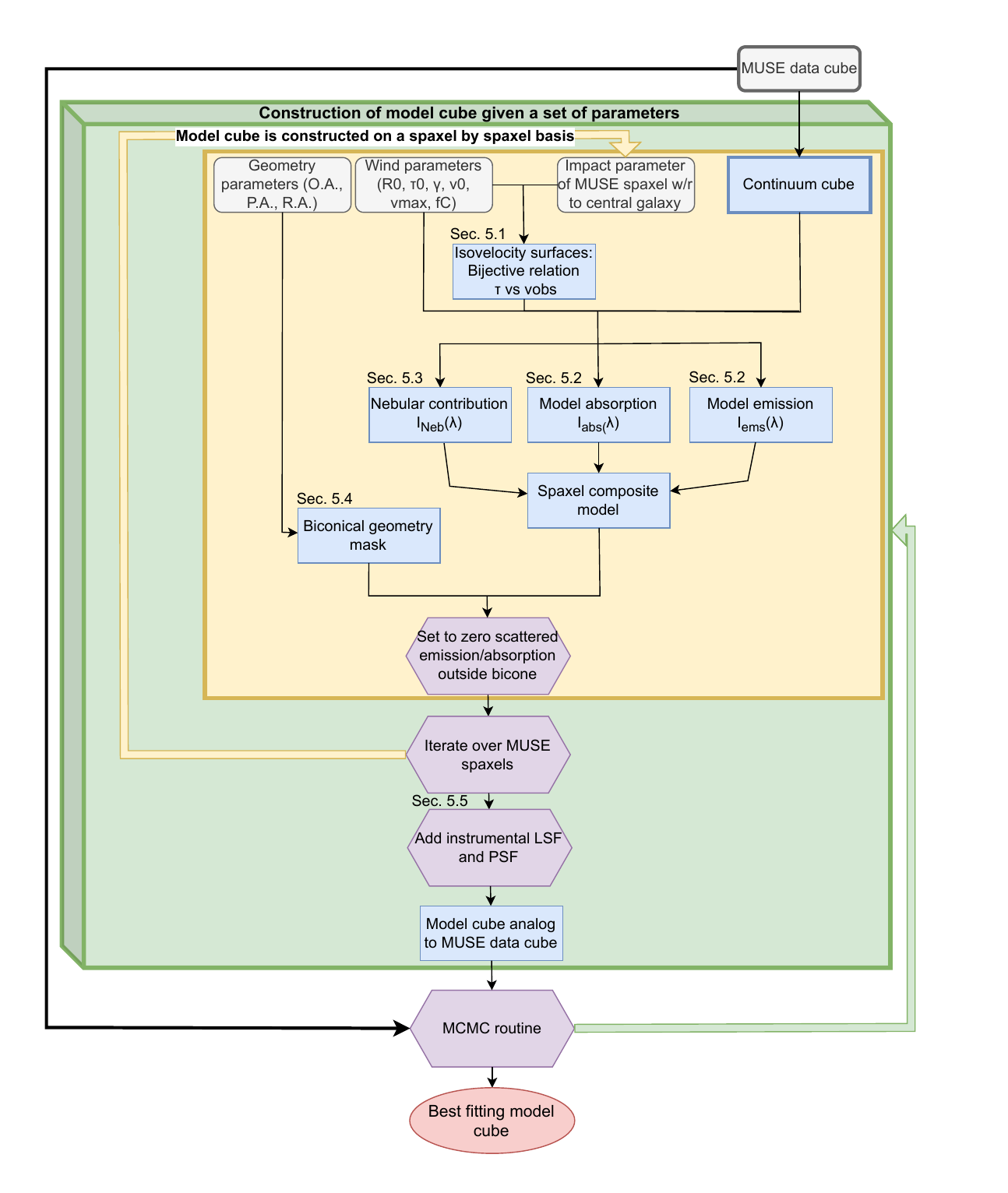}
\caption{Flow chart summarizing our modeling and fitting scheme. Gray rounded boxes correspond to inputs for the model, blue rectangles correspond to intermediate outputs, and purple hexagons correspond to processes. The final output is indicated with a red ellipse. For a given set of parameters, a model cube is constructed on a spaxel-by-spaxel basis. The iteration over different individual spaxels is indicated with the yellow area. We used an MCMC routine to find the best-fitting parameters that match our data cube. The complete MCMC iteration over the parameter space is enclosed within the green (including the yellow) area. The inputs and intermediate outputs surrounded by a thick box are static, that is, they do not change over any of the iterations performed. In the top-left part of some boxes, we indicate the section where the concept inside the box is described. See Secs.~\ref{sec:model} and~\ref{sec:fitting_method} for a comprehensive description of our modeling and fitting approach.}
\label{fig:flow_chart}
\end{figure*}

The main goal of this paper is to design a semianalytic model, with only a few parameters to avoid overfitting, that can reproduce key features observed of the \ion{Mg}{II} halo of \#884, in terms of the spatial distribution and line profiles of the \ion{Mg}{II} emission and absorption, and that can be fitted to spatially resolved IFU observations. In the following subsections, we describe in detail the different components and other relevant aspects considered for our model, in order to achieve this goal. Figure~\ref{fig:flow_chart} provides a flow chart that summarizes our modeling approach. Further in the paper, we describe the fit of our model to the data, aiming to unveil some of the key physics at play in the circumgalactic gas.

\subsection{Basic model}
\label{sec:basic_model}

\begin{figure*}
\includegraphics[width=\textwidth]{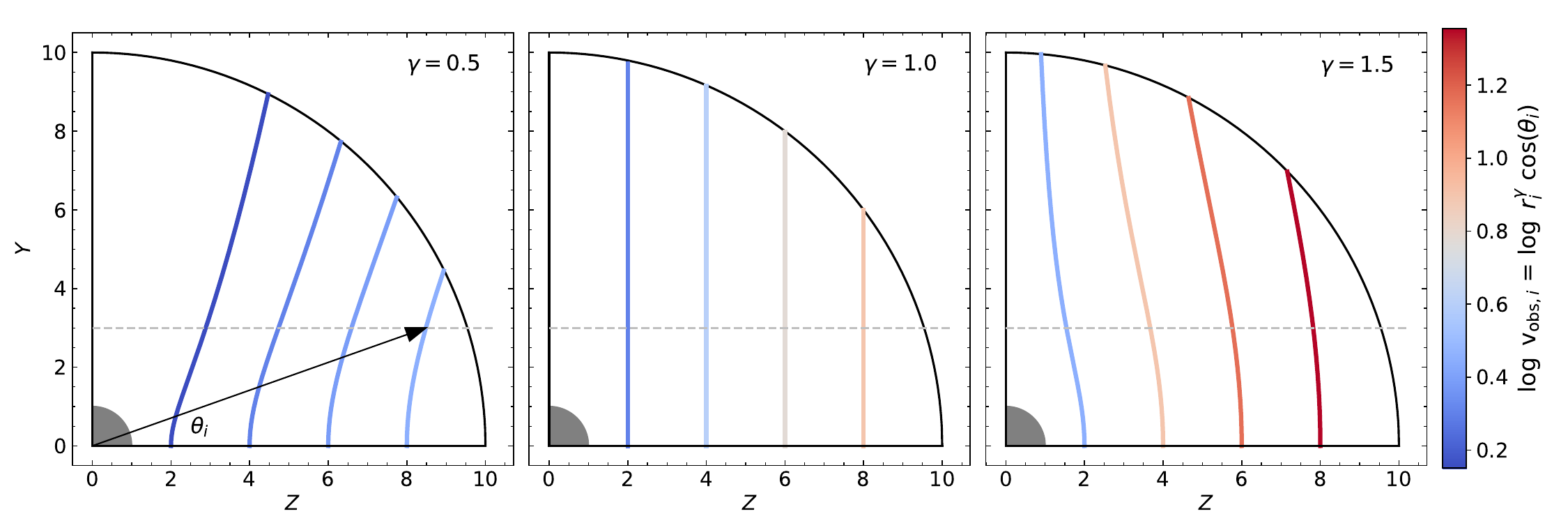}
\caption{Isovelocity curves in the spherical outflow model. The curves were calculated following Eq.~\ref{eq:radius_velocity} for three different values of $\gamma$, setting $R_0$ and $v_0$ to 1. The observer sits far away from the object toward the positive z-direction. The gray dashed line represents an arbitrary line of sight. Due to projection effects, the observed velocity of each spherical shell ($v_{\mathrm{obs}}$) corresponds to the velocity component parallel to the line of sight, and will be smaller than its intrinsic velocity (except for an impact parameter of zero, in which case they are the same). Thus, $v_{\mathrm{obs}}$ for a given spherical shell depends on the impact parameter of the line of sight, and the angle $\theta$ between the intersection of the spherical shell with the line of sight, and the $z$-axis (i.e., a higher $\theta$ implies that a lower fraction of the intrinsic radial velocity is parallel to the line of sight). The isovelocity surfaces are defined as the set of points within the outflow for which the observed velocity is identical, due to projection effects. Different values of $\gamma$ lead to different shapes of the isovelocity curves. In particular, $\gamma = 1$ leads to perfectly vertical surfaces, meaning that the increase in radius (and thus, velocity) by moving upward in the $y$ direction is exactly compensated by the decrease in $\cos(\theta)$, that is, a smaller fraction of the intrinsic velocity of the spherical shell is parallel to the line of sight, causing the projected velocity to remain unchanged. The most relevant aspect is that the isovelocity curves do not cross each other, implying that any line of sight intersects each isovelocity curve only once. Ultimately, the consequence of this is that for any arbitrary line of sight, different observed velocities map specific unique locations within the outflow (in the $z$ direction). This bijective relation between the velocity and physical space allows us to probe the outflow properties on a 3-D basis provided that we have sight lines for different locations in the $x-y$ plane (i.e., integral field data).}
\label{fig:sobolev}
\end{figure*}

In this section, we describe the outflow model introduced by \citet{Scuderi1992} in the context of stellar winds and further developed by \citet{Scarlata2015}, \citet{Carr:2018hc}, and \citet{Carr2019} to model the scattered resonant emission in expanding envelopes around galaxies. We also explain how we adapt this modeling scheme to integral field spectroscopic data. The basic model consists of a spherically symmetric source of radius $R_{0}$ isotropically emitting continuum radiation, surrounded by a spherical envelope of gas (wind) expanding from radius $R_{0}$ up to a terminal radius $R_{\rm max}$. The velocity at which the spherical envelope expands ($v$) increases with radius ($r$), according to a power law of exponent $\gamma$ such that

\begin{equation}
\label{eq:vel_field}
\begin{aligned}
    &v = v_{0} \, \biggl(\frac{r}{R_{0}} \biggr)^{\gamma} &\mathrm{\;for\;} r < R_{\rm max} \\
    &v =v_{\rm max}   &\mathrm{\;for\;} r \geq R_{\rm max},\\
\end{aligned}
\end{equation}

\noindent where $v_{0}$ corresponds to the initial velocity of the wind (i.e., at $R_{0}$), and $v_{\mathrm{max}}$ is the terminal velocity of the wind at $R_{\mathrm{max}}$. A wind velocity that increases with radius is consistent with results from simulations \citep[see, e.g.,][]{Cooper2008}, and it is also consistent with the down-the-barrel observations reported by \citet{Martin:2009fl}, although different, more complex velocity profiles are possible \citep[see, e.g.,][]{Murray2011}. The density profile corresponding to Eq.~\ref{eq:vel_field}, assuming mass conservation through the outflow, is

\begin{equation}
\label{eq:density}
    n(r) = n_{0} \, \biggl( \frac{R_{0}}{r}  \biggr)^{\gamma + 2},
\end{equation}

\noindent where $n_{0}$ is the gas density at $r = R_{0}$. Provided the velocity gradient in Eq.~\ref{eq:vel_field} is large, the photons produced by the central source will interact with the outflowing material only at the specific radius where the absorbing ions are at resonance (due to their Doppler shift), an assumption commonly known as the "Sobolev" approximation. However, due to projection effects, the line-of-sight velocity $v_{\rm obs}$ measured by an observer for a given spherical shell at radius $r$ will differ from its real velocity, and will depend on the projected distance ($\epsilon$) of the line of sight to the central source (a.k.a., impact parameter), going from $v(r)$ at the center ($\epsilon = 0$), to zero at $\epsilon = r$ \citep[see also][ for a more detailed explanation]{Scarlata2015, Carr:2018hc}. On the other hand, it can be shown that the surfaces of constant $v_{\rm obs}$ do not cross each other, and hence, for any given line of sight, there is a bijective relation between frequency (or velocity) space and spatial location within the expanding envelope (because any sight line intersects different isovelocity surfaces only once). Figure~\ref{fig:sobolev} shows the shape of the (observed) isovelocity curves for $\gamma$ = 0.5, 1.0, and 1.5.

Specifically, for a line of sight at an arbitrary impact parameter $\epsilon$, the position $r$ within the envelope that would correspond to an observed velocity $v_{\rm obs}$ is determined by the expression

\begin{equation}
    \label{eq:radius_velocity}
    r^{2} - R_{0}^{2} \, \biggl( \frac{v_{\rm obs}}{v_{0}} \biggr)^{2/\gamma} \biggl(1 - \biggl[\frac{\epsilon}{r}\biggr]^{2} \biggr)^{\frac{\gamma - 1}{\gamma}} - \epsilon^{2}= 0.
\end{equation}

\noindent Thus, the optical depth ($\tau$) can be evaluated for each wavelength on the corresponding interaction surface as a function of the physical properties of the wind and atomic constants following \citet{Castor1970}:

\begin{equation}
    \label{eq:definition_tau}
    \tau (r) = \frac{\pi e^{2}}{mc} \, f_{lu} \, \lambda_{lu} \, n_{l}(r) \, \biggl[ 1 - \frac{n_{u} \, g_{l}}{n_{l} \, g_{u}} \biggr] \frac{r/v}{1 + \sigma \cos^{2} \phi },
\end{equation}

\noindent where $\phi$ is the angle between the velocity and the trajectory of the photon, $f_{ul}$ and $\lambda_{ul}$ correspond to the oscillator strength and wavelength for the $ul$ transition, respectively, and $\sigma = \frac{d \ln (v)}{d \ln (r)} - 1$. Assuming the velocity law in Eq.~\ref{eq:vel_field} and neglecting stimulated emission (i.e., $\biggl[ 1 - \frac{n_{u} \, g_{l}}{n_{l} \, g_{u}} \biggr] = 1$), we get that the optical depth can be expressed as
\begin{equation}
 \tau (r) = \frac{\pi e^{2}}{mc} \, f_{lu} \, \lambda_{lu} \, n_{0} \biggl( \frac{R_{0}}{r}  \biggr)^{\gamma + 2} \frac{r/v}{1 + (\gamma - 1) \cos^{2} \phi }.
\end{equation}

\noindent Finally, defining
\begin{equation}
\label{eq:tau_0_n_0}
    \tau_{0} = \frac{\pi e^{2}}{mc} \, f_{lu} \, \lambda_{lu} \, n_{0} \, \frac{R_{0}}{v_{0}},
\end{equation}
we can write the optical depth as
\begin{equation}
    \label{eq:optical_depth}
    \tau (r) = \frac{ \tau_{0}}{1 + (\gamma - 1) \, \cos^{2} \phi } \biggl( \frac{R_{0}}{r}  \biggr)^{2\gamma + 1}.
\end{equation}

\noindent Altogether, eqs.~\ref{eq:radius_velocity} and ~\ref{eq:optical_depth} provide a bijective relation between observed velocity and radius of resonance, that is, the radius where continuum photons will interact with the outflowing ions, given their Doppler shift (for a given sight line, defined by its impact parameter), and the value of optical depth at any radius within the outflow, respectively. Combining these two equations it is possible to calculate $\tau$ as a function of $v_{\rm obs}$ (and, therefore, wavelength $\lambda$).

\subsection{Modeling integral field spectroscopic data}
Our goal is to produce a model cube that matches the data, for a given set of parameters. Thus, we compute the modeled absorption and emission profiles independently for each discrete sight line (i.e., MUSE spaxel), according to their respective impact parameter to the central source (also taking into account instrumental effects; see Sect.~\ref{sec:psf_lsf}). 

For a given MUSE spaxel, the fraction of the continuum flux absorbed at each wavelength by the spherical envelope located at resonance with the photons emitted by the central source at wavelength $\lambda$ is given by

\begin{equation}
    I_{\rm abs}(\lambda) = 1 - e^{-\tau(\lambda)}.
\end{equation}

We note that for a high optical depth, $I_{\rm abs} \approx 1$, meaning that the continuum emission is completely absorbed. 
\\
Resonant photons absorbed by the spherical envelope can then be re-emitted toward the observer. We calculate the emission produced by the gas in the envelope by assuming that the re-emitted photons are able to escape their resonant shell without further interactions (i.e., single scattering approximation). The profile of the emission for any given sight line, as a function of wavelength, can be then expressed in units of the continuum radiation as

\begin{equation}
\label{eq:emission_part}
    I_{\rm ems}(\lambda) = \frac{1 - e^{-\tau (\lambda)}}{4 \pi r_{\lambda}^{2}},
\end{equation}

\noindent where $r_{\lambda}$ is the radius of resonance for the photons emitted at wavelength $\lambda$, calculated following Eq.~\ref{eq:radius_velocity}. For high values of $\tau$, $I_{\rm ems} \approx 1/4 \pi r_{\lambda}^{2}$, that is, the absorbed photons are re-emitted in a spherical shell of radius $r_{\lambda}$. Integrating the flux over this sphere adds up to the entirety of photons absorbed. Thus, for each MUSE spaxel, $I_{\mathrm{ems}}$($\lambda$) must be rescaled by the area subtended by the spaxel projected onto a sphere of radius $r_{\lambda}$.

However, we note that while the absorbing component is scaled by the level of continuum in any specific sight line, the emission contribution must be scaled to the total continuum radiation emitted by the central source. This is because while the absorption is only meaningful relative to a continuum, the emission component does not depend on the presence of a continuum in the same sight line, but it is powered by the total production of continuum photons by the central source. 

By construction, the absorption component is confined to wavelengths that correspond to $v_{\rm obs} < 0$, since only the part of the spherical envelope moving toward the observer, between the observer and the central source, is able to absorb continuum radiation in the line of sight. On the other hand, both the blueshifted and redshifted parts of any given spherical shell are able to scatter photons into the line of sight. However, we note that since $R_{0}$ represents the launching radius of the wind, wavelengths that correspond to a radius smaller than $R_{0}$ do not absorb or emit scattered radiation.
\\

The resulting P-Cygni profile of the outflow can be computed for any specific sight line as

\begin{equation}
    I(\lambda) = I_{\rm cont, local} (1 - I_{\rm abs}) + I_{\rm cont, total}\;I_{\rm ems},
\end{equation}
where $I_{\rm cont, local}$ and $I_{\rm cont, total}$ correspond to the local level of continuum in the sight line, and the total continuum of the galaxy in the corresponding wavelengths, respectively, with $R_0$, $\tau_{0}$, $\gamma$, $v_{0}$, and $v_{\rm max}$ being free parameters of the model. As explained in Sect.~\ref{sec:nb}, the level of the continuum in each spaxel is calculated with a running median in the spectral direction of the data cube, within a window of $100$ pixels. This continuum is then passed to the model cube. Thus, by construction, the continuum in the data cube and model cube are identical. 

Finally, due to our inability to resolve the half-light radius of the central galaxy (and thus, spatial variations in the spectral profile of the absorption component), we assume that all the observed continuum light is originally emitted from the central half-light radius. Hence, we preserve the same spectral profile of the absorption $I_{\mathrm{abs}}$ across the modeled FoV, and in each spaxel it is scaled to the local continuum $I_{\mathrm{cont,local}}$.

To avoid including unphysical \ion{Mg}{II} absorption far away from the central galaxy, where the local level of continuum might be different from zero, either due to noise effects or due to scattered light from nearby (in projected distance) sources, we have set a maximum radius for absorption, beyond which the absorption is set to zero, even if the continuum is different from zero. This maximum absorption radius has been empirically determined as the radius where the radial profile computed in the absorption-dominated band of the P-Cygni profile (see Sect.~\ref{sec:radial_profiles} for details on how the absorption-dominated band is defined) intersects the continuum. That is, the radius beyond which there is no more \ion{Mg}{II} absorption in the data cube. We find this radius to be $1\farcs1$, which corresponds to a physical distance of $\sim8.2$ kpc at $z=0.737$.

\subsection{Nebular emission contribution}

Observations of intermediate-redshift $(0.2 \leq z \leq 2.3)$ star-forming galaxies have shown that massive galaxies typically show net \ion{Mg}{II} absorption, whereas \ion{Mg}{II} emitters are quite abundant among lower mass ($\log (M_\star$/M$_\odot) \lesssim 9.5$) and lower metallicity ($7.5 \lesssim 12 + \log \mathrm{O/H} \lesssim 8.4$) objects \citep{Weiner:2009cf,Erb2012,Guseva:2013jp,Finley:2017eg,Henry:2018gd,Feltre:2018in}. Several of these authors demonstrated that photoionization models of \ion{H}{II} regions indeed predict appreciable nebular \ion{Mg}{II} emission, of similar levels as observed. We note that our galaxy \#884 does not meet the criteria for typical \ion{Mg}{II} emitters as those that were investigated by the above authors, because its spectrum is strongly absorption-dominated in the central arcsec$^2$. Nevertheless, \#884 is an actively star-forming galaxy with presumably a lot of \ion{H}{II} regions, so it could produce nebular \ion{Mg}{II} emission, of which a fraction might be able to escape directly without prior scattering in the outflow. 

Thus, we modeled this possible extra source of \ion{Mg}{II} photons in a given sight line as a Gaussian (centered at $0$ km s$^{-1}$) whose amplitude is proportional to the local stellar continuum, through a free parameter $f_{C}$. Specifically, we define

\begin{equation}
    I_{\rm Neb}(\lambda) = (f_{C}\times I_{\rm cont, local}) e^{-\frac{1}{2}\frac{(\lambda - \lambda_{0})^{2}}{\sigma_{\rm Neb}^{2}}},
\end{equation}

\noindent where the standard deviation of the nebular emission $\sigma_{\rm Neb}$ is fixed to the value of velocity dispersion measured for H$\beta$ ($\sim 45$ km s$^{-1}$), another strong, available non-resonant nebular emission line. Lastly, the intrinsic emissivity ratio of nebular \ion{Mg}{II} $\lambda$2796 to \ion{Mg}{II} $\lambda$2803 emission is set by atomic physics to be two \citep[see, e.g.,][]{Chisholm2020}, so we fix this ratio in our model by using half of the amplitude for the \ion{Mg}{II} $\lambda$2803 line.

This additional component does not take into account possible in-place production of \ion{Mg}{II} photons in the halo. This possibility is described in detail in \citet{Zabl2021}. Although as described earlier in Sect.~\ref{sec:nb}, the emission line ratios in the spectrum of \#884 suggest that the extended \ion{Mg}{II} emission primarily originates from the resonant scattering of continuum photons, we cannot rule out a partial contribution of non-scattering photons (e.g., shocks), especially given the presence of extended emission from optical non-resonant lines (Wisotzki et al. in prep.).

\subsection{Biconical geometry}
\label{sec:biconic_geometry}
Observational evidence suggests that the circumgalactic medium is not isotropically distributed around a central galaxy \citep[e.g.,][]{Schroetter:2019es}. Moreover, the data are consistent with a picture of \ion{Mg}{II} being predominantly present in outflow cones and extended disc-like structures.  \citep[e.g.,][]{Bouche:2012ho,Schroetter:2019es,Zabl:2019ija,FernandezFigueroa2022, Guo2023b}. Thus, a strictly spherical geometry represents an oversimplification of the problem. We introduce a biconical geometry into the model, where the scattered emission and absorption at a given wavelength are set to zero if the wavelength corresponds to a physical location outside of the outflow cones. \citet{Carr:2018hc}  have also explored the effect of a biconical geometry in the integrated spectra of galactic outflows produced by their Semi-analytical Line Transfer \citep[SALT,][]{Scarlata2015}.

\begin{figure}
\includegraphics[width=\columnwidth]{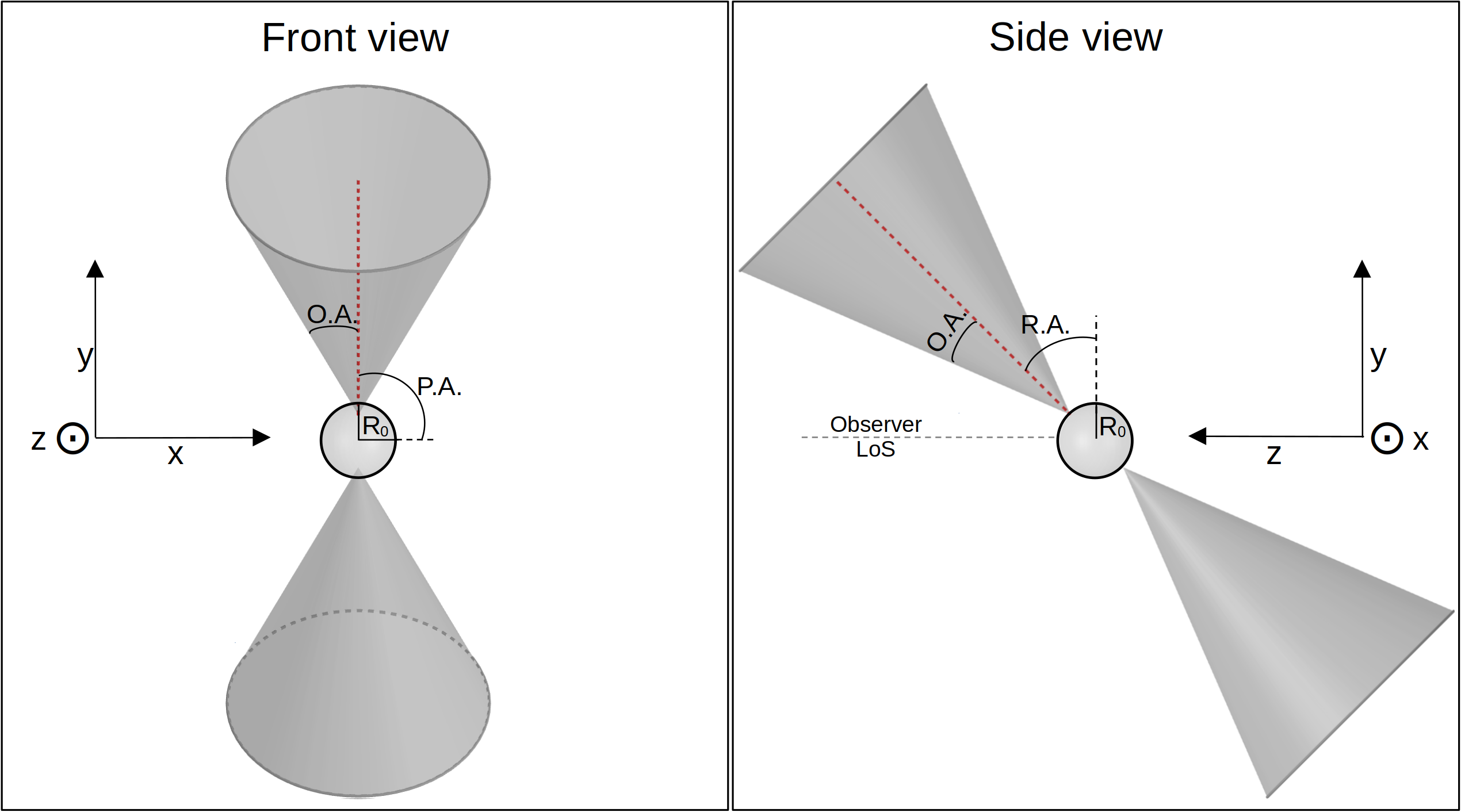}
\caption{Schematic representation of the outflow geometry parametrization described in Sect.~\ref{sec:biconic_geometry}. The figure illustrates the front (\textit{left}) and side (\textit{right}) view of an outflow characterized by an arbitrary O.A.,  P.A. $= 90^{\circ}$ and R.A. $= 45^{\circ}$. The opening angle represents the angle between the axis and the surface of each outflow cone, the position angle corresponds to the angle of the outflow with respect to the horizontal axis, in the plane of the sky, and the rotation angle measures the rotation of the outflow in the plane defined by $x = 0$, with respect to the $y$-axis. The red line shows the cone axis. The observer is located at a large positive $z$ value.}
\label{fig:bicone}
\end{figure}

The biconical geometry can be fully described through three additional parameters; an opening angle (O.A.), a position angle (P.A.), and a rotation angle (R.A.). Figure~\ref{fig:bicone} provides a schematic representation of the angles that describe this parametrization. The opening angle represents the angle between the central axis and the surface of each cone. It can take values in the range of $0 < \mathrm{O.A.} \leq  \frac{\pi}{2}$ radians, where $0$ radians would correspond to a closed cone (i.e., no outflow), and $\frac{\pi}{2}$ radians is the case for a totally spherical outflow. That is, we recover the spherical outflow scenario as a limiting case in this parametrization.

The position angle corresponds to the angle of the outflow with respect to the horizontal axis, in the plane of the sky (i.e., perpendicular to the line of sight; see the left panel of Fig.~\ref{fig:bicone}). It is defined within the range $0 \leq \mathrm{P.A.} \leq \pi$, where a P.A. of zero represents a completely horizontal outflow in the plane of the sky, and a P.A. of $\frac{\pi}{2}$ corresponds to the vertical case.

Finally, the rotation angle measures the rotation of the outflow in the plane defined by $x = 0$, with respect to the $y$-axis. That is, a rotation perpendicular to the plane of the sky, toward the observer (see the right panel of Fig.~\ref{fig:bicone}). It is defined in the range $-\frac{\pi}{2} \leq \mathrm{R.A.} \leq \frac{\pi}{2}$, where zero represents an outflow in the plane of the sky, perpendicular to the line of sight, and both $-\frac{\pi}{2}$ and $\frac{\pi}{2}$ correspond to the case where the outflow is pointing directly to the observer.

Hence, for a given set of O.A., P.A., and R.A., we set the scattered emission outside the biconical volume to zero, whereas inside the outflow it is given by Eq.~\ref{eq:emission_part}

\subsection{Line spread function and point spread function convolution}
\label{sec:psf_lsf}

Accounting for the instrumental effects on the spectral line profile and on the spatial light distribution is essential for a meaningful comparison between our model, described in the previous subsections, and the data. These nonphysical effects are encoded by the instrumental LSF and PSF, where the first describes the spectral broadening of a given spectral line due to instrumental effects, and the second characterizes the spatial broadening of the light distribution of a point-like source in the detector. Both the LSF and PSF of the MUSE-HUDF are well characterized in \citet{Bacon:2017hn}, and we use the same parametrization to model these effects. Specifically, we apply an equivalent spectral and spatial convolution to the modeled data cube before comparing it with the observed data cube. Here, we provide a brief summary of our approach to account for these effects, but we refer the reader to \citet{Bacon:2017hn} for a more detailed description.

\subsubsection{Line spread function}
The LSF is modeled as a Gaussian function, whose width was measured by fitting a Gaussian to a set of sky lines, across all spaxels in the mosaic. \citet{Bacon:2017hn} find that the FWHM of the LSF varies smoothly with wavelength, as a second-degree polynomial:

\begin{equation}
\mathrm{FWHM}(\lambda) = 5.835 \times 10^{-8} \lambda^{2} - 9.080 \times 10^{-4} \lambda + 5.983,
\end{equation}

\noindent where $\lambda$ and FWHM are both in $\AA$. For the observed wavelength of \ion{Mg}{II}, this represents a FWHM of $\sim 2.96$ $\AA$ for the LSF.

\subsubsection{Point spread function}
During commissioning, a detailed analysis of the MUSE PSF showed that it was very well modeled by a Moffat circular function with constant $\beta$ and a FWHM that varies linearly with wavelength. The values of the FWHM and $\beta$ were determined by matching the resolution of a set of HST broadband images, and analog synthetic images constructed from the MUSE data cubes, using the same spectral response and total flux calibration as for the HST images, provided that the HST PSF is well known\footnote{This methodology was adopted because most of MUSE fields do not have bright stars to model the PSF independently.}.
\citet{Bacon:2017hn} find that the PSF is well represented by a Moffat with $\beta = 2.8$ and FWHM that decreases with wavelength, from $0\farcs71$ at the blue end, to $0\farcs57$ at the red end. The observed wavelength of the \ion{Mg}{II} transition is very close to the blue end of the MUSE wavelength range ($\sim 4860$ $\AA$), which corresponds to a FWHM of $\sim 0\farcs7$ for the PSF.

\section{Fitting method}
\label{sec:fitting_method}
\subsection{Modeled signal}
We use the model described in the previous section to reproduce the \ion{Mg}{II} P-Cygni profile exhibited by the galaxy MUSE-HUDF $\#884$. To do this, we consider the wavelength interval of 2790 - 2809 $\AA$  (rest-frame), thus modeling both \ion{Mg}{II} transitions simultaneously (see Fig~\ref{fig:aperspec1}). In order to avoid contamination from nearby galaxies (in projected distance, particularly from BG1 and FG2; see Sect.~\ref{sec:discenv}), we fit the \ion{Mg}{II} emission profile in a circular FoV with a radius of $3\farcs5$ ($\sim$26 kpc) around the central galaxy (blue aperture in Fig~\ref{fig:mg2nb1}).

We model the \ion{Mg}{II} $\lambda \lambda 2796$, $2803$ transitions at their corresponding rest-frame velocities, and superpose their respective absorption and emission contributions on the same spectrum. We use the same set of parameters to model both transitions, except for $\tau_{0}$. Since the oscillator strength of \ion{Mg}{II} $\lambda 2796$ is essentially two times larger than that of \ion{Mg}{II} $\lambda 2803$ \citep{Kelleher2008}, we use an optical depth $\tau = \tau_{0}/2$ for \ion{Mg}{II} $\lambda 2803$, as the optical depth depends linearly on the oscillator strength (see Eq.~\ref{eq:definition_tau}).

Hence, for each MUSE spaxel, given a set of parameters and a continuum level, we generate a composed model spectrum of the \ion{Mg}{II} $\lambda \lambda 2796$, $2803$ doublet, resulting in a model cube as the final product.

\subsection{Description of the fitting}

The best-fit parameters and associated uncertainties in our fitting procedure were derived using a Markov chain Monte Carlo (MCMC) analysis implemented in the {\sc emcee} python package \citep{ForemanMackey:2013io}.  For each MCMC step, we compute a model cube and the respective likelihood of the used set of parameters. 

We used the following priors for the outflow parameters, described in Sect.~\ref{sec:basic_model}:
\begin{gather}
R_{0}\,\,\,\, \sim \mathcal{N}(1.5,0.2) \\
\tau_0\,\,\,\,\, \sim \mathcal{L}\mathcal{U}(0,4) \\
\gamma \,\,\,\,\,\,\, \sim \mathcal{U}(0.2,1.0) \\
v_0 \,\,\,\,\, \sim \mathcal{U}(20,120) \\
v_{\mathrm{max}} \sim \mathcal{U}(300,550) \\
f_{C} \,\,\, \sim \mathcal{L}\mathcal{U}(-3.0,0.5),
\end{gather}
where  $\mathcal{N}(\mu,\sigma)$ stands for a normal distribution with mean $\mu$ and standard deviation $\sigma$. $\mathcal{L}\mathcal{U}(a,b)$ and $\mathcal{U}(a,b)$ stand for log-uniform and uniform distributions, within the $(a,b)$ intervals, respectively. 

We note that for $R_{0}$ we chose a normal prior rather than a uniform one. This is because we know the approximate size of the galaxy from HST photometry (see Sect.~\ref{tab:global}), and this value should correspond approximately to $R_{0}$ (only approximately because $R_0$ represents a simplification of the model, assuming that the continuum central source is spherical). Thus, we exclude scenarios where $R_0$ is much larger or much smaller than the actual size of \#884. Furthermore, given the MUSE spatial and spectral resolution, $R_{0}$ is only partially constrained, so we purposely use narrow priors around the half-light radius measurement, instead of using a fixed $R_0$ value.

We have also included an extra parameter $\Delta v$ to account for any kinematic offset between the systemic velocity and the P-Cygni emission. Its prior is given by
\begin{equation}
\Delta v \sim \mathcal{U}(80,130) \\,
\end{equation}
and it has units of kilometers per second. 

For the geometric parameters, described in Sect.~\ref{sec:biconic_geometry}, we considered the following priors:

\begin{gather}
\mathrm{O.A.} \sim \mathcal{U}(40,70) \\
\mathrm{P.A.} \sim \mathcal{N}(120,10) \\
\mathrm{R.A.} \sim \mathcal{U}(40,90)
\end{gather}
 
\noindent in units of degrees. We note that for P.A. we also chose a normal prior, instead of a more agnostic uniform distribution. This choice is justified on the basis that the \ion{Mg}{II} pseudo-narrowband image extracted from the MUSE data cube (see Fig.~\ref{fig:mg2nb1}) shows a preferred direction on the spatial distribution of the \ion{Mg}{II} emission, along the diagonal axis, from the top-left to the bottom-right corner of the image.

Finally, we included an additional term $f$ to account for the existence of additional variance in the data. The extra variance ($\sigma$) is proportional to the model such that

\begin{equation}
    \sigma^{2} =  f^{2} M^{2},
\end{equation}
where $M$ is the value of the model, evaluated in the independent variable, given a set of parameters. The prior on $f$ is given by
\begin{equation}
f \sim \mathcal{L}\mathcal{U}(-5,-1).    
\end{equation}

 The uniform priors used are designed to be agnostic, broad enough to sample the posterior distribution of the different parameters, but also not including regions of the parameter space that numerous previous runs of the fitting showed are far away from the higher likelihood region. We use 22 Markov chain Monte Carlo (MCMC) chains to sample the posterior distributions, each with 6000 iterations. We discard 500 burn-in iterations before convergence for each chain. The convergence of the MCMC chains is ensured by using the $\hat{R} \approx 1$ criterion \citep{Gelman1992}. Essentially, $\hat{R} \approx 1$ implies that the between-chain variance and the within-chain variance are approximately the same.
\section{Fitting results}
\label{sec:fitting_results}
\subsection{Best-fitting model}
\label{sec:best_fitting_model}
\begin{figure*}
\includegraphics[width=\textwidth]{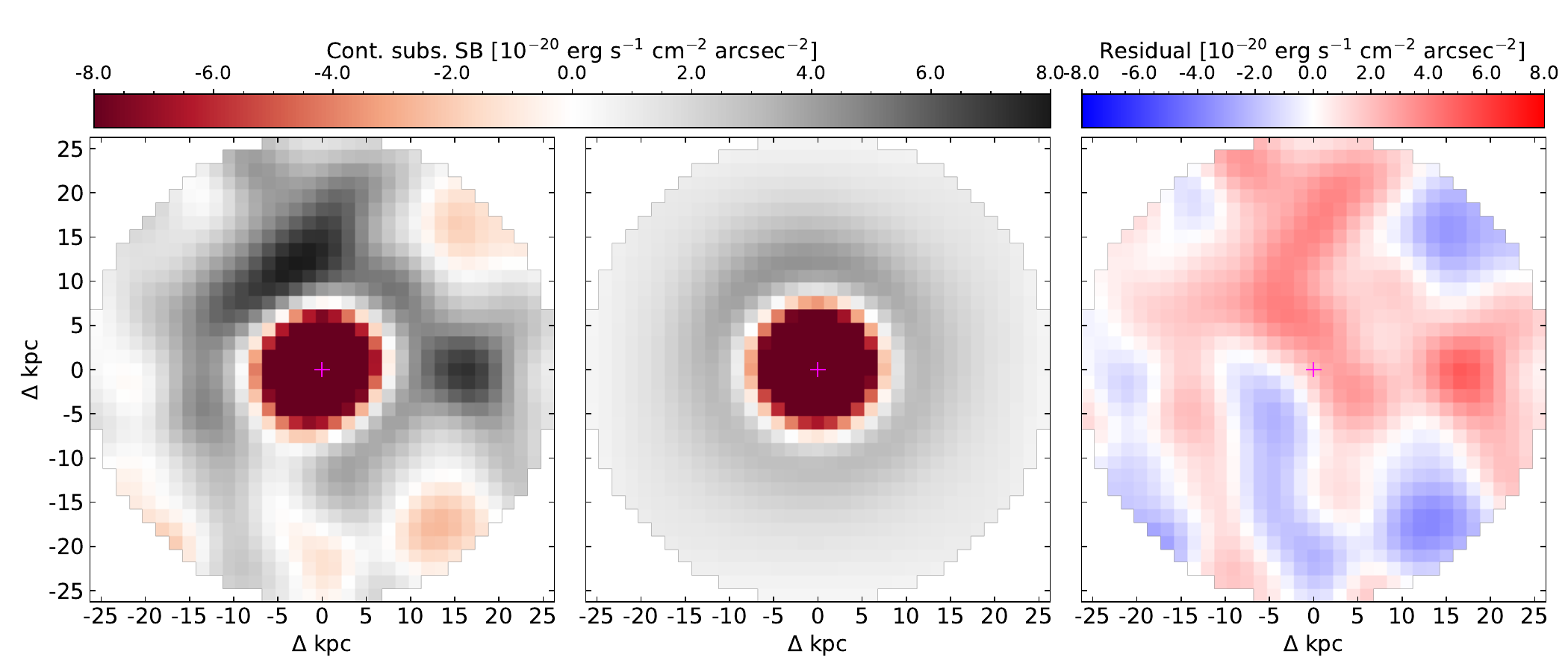}
\caption{Comparison of the pseudo-narrowband images convolved with a Gaussian of 1\arcsec\ FWHM, computed for the MUSE data cube (\textit{left}) and the best-fitting model cube (\textit{center}), collapsing the \ion{Mg}{II} wavelength range, after subtracting the continuum from both cubes, for the full modeled FoV. The residuals are shown in the right panel. The magenta cross indicates the center of the FoV for reference. There is a good qualitative agreement between the images, with each displaying an absorption-dominated center, shown in red, and a more spatially extended halo of emission \ion{Mg}{II} emission, shown in gray colors. However, there are also clear differences in the morphology and strength of the \ion{Mg}{II} emission.}
\label{fig:cont_subs_FoV}
\end{figure*}

In this section, we investigate the best-fitting model obtained using the fitting routine described in the previous section. Figure~\ref{fig:cont_subs_FoV} shows a pseudo-narrowband image convolved with a Gaussian of 1\arcsec\ FWHM, for the MUSE data cube and the best-fitting model cube, collapsing the \ion{Mg}{II} wavelength range, for the FoV where we compute the model (in the case of the model cube, this convolution is applied in addition to the MUSE PSF such that both cubes are similarly convolved). The stellar continuum has been subtracted from both images to better compare the morphology of the \ion{Mg}{II} absorption and emission components.

Qualitatively, the model reproduces the shape of the \ion{Mg}{II} emission and absorption structure. In both images, the central region is strongly dominated by \ion{Mg}{II} absorption, while the emission is preferentially located further away from the center, although there are some clear differences in the morphology of the \ion{Mg}{II} emission. The apparent absorption features in the data at the edge of the image are either a consequence of noise, or contaminant continuum emission from a background source. In the following,  we assess more quantitatively the ability of the model to reproduce different aspects of the data. Additionally, in Appendix~\ref{sec:aperture_integrated}, for comparison purposes, we show the best-fitting model when only the aperture-integrated spectrum is modeled, rather than performing a spatially resolved fitting.

\begin{figure*}
\includegraphics[width=\textwidth]{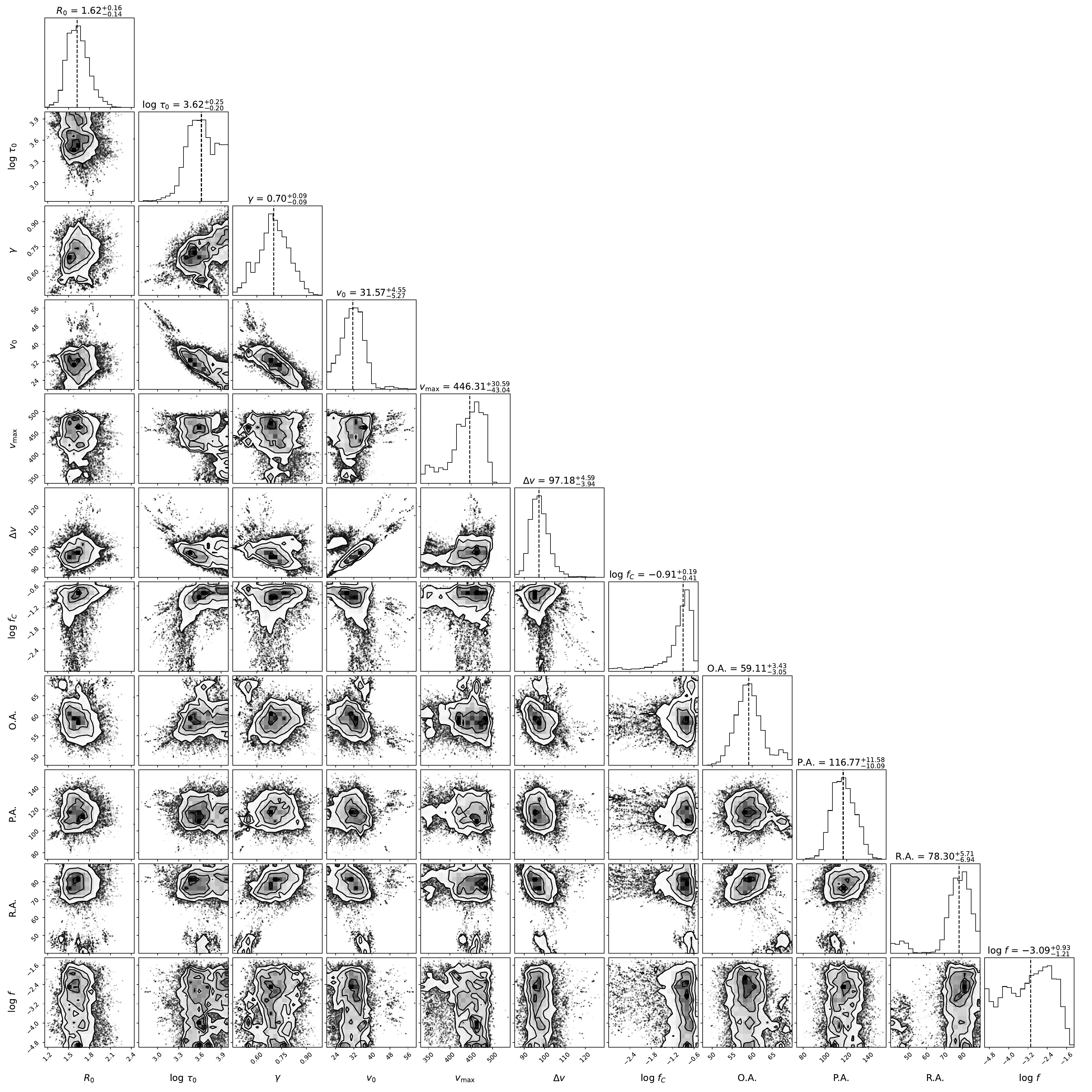}
\caption{Posterior distributions of each one of the model parameters described in Sect.~\ref{sec:model} and their respective covariance. The median value of each posterior distribution is indicated in the top panels. The uncertainties show the percentiles 32 and 68 of the distribution.}
\label{fig:posterior_dist}
\end{figure*}

Figure~\ref{fig:posterior_dist} shows the posterior distributions of each of the model parameters described in Sect.~\ref{sec:model}, and their respective covariance.

Despite the inability of MUSE to spatially resolve $R_0$, the posterior distribution of $R_{0}$ is not centered around the prior value of 1.5 kpc. Instead, the posterior distribution is shifted toward a slightly larger radii ($R_{0} = 1.6\pm 0.2$ kpc), showing that a continuum source size larger than 1.5 kpc is marginally favored to reproduce the data (although 1.5 kpc is within the error bars of the best-fitting value). However, we acknowledge that the width of this posterior distribution is partially driven by the narrow prior selection and that $R_{0}$ is only marginally constrained in our dataset. 

On the other hand, the posterior distribution of the optical depth $\tau_{0}$ exhibits a clear maximum at $\tau_{0} = 4000\pm2000$ (albeit with a tail toward even higher values, hence, the large uncertainty in the parameter). It is remarkable that despite the heavily saturated nature of the \ion{Mg}{II} absorption line, our modeling scheme still allows us to put a constraint on the gas optical depth. This is possible because although the absorption is completely saturated in the inner part of the outflow, our model is still sensitive to the changes in the optical depth toward larger radii, that is, the radial gradient of $\tau$. As a consequence, we find some degree of covariance between $\tau_0$ and the quantities that determine the velocity (and $\tau$) gradient, $\gamma$ and $v_{0}$. The combination of parameters of our best-fitting model leads to $\tau = 1$ only at $R \sim 60$ kpc. In principle, beyond this radius, we would not need the Sobolev approximation since most photons would simply escape the galaxy without scattering with the ions in the CGM, independently of the velocity gradient. However, this only occurs at radii much larger than the impact parameters considered in our model, and therefore, it does not represent a conflict with our assumptions.

We also note that the blue wing of the \ion{Mg}{II} absorption in the innermost aperture shown in Fig.~\ref{fig:aperspec1} is particularly sensitive to the radial gradient of $\tau$, since it provides a down-the-barrel view of the outflow, where solely changes in $\tau$ along the line of sight determine its spectral shape. In Appendix~\ref{sec:inner_aperture_modeling}, we explore how the inferred optical depth changes when we only model this inner region.

The kinematics of the gas, parameterized by $\gamma$, $v_{0}$ and $v_{\mathrm{max}}$ are generally well constrained, exhibiting a posterior distribution with a clear maximum, and a relative error $\lesssim 15\%$. The radial acceleration of the expansion is given by $\gamma = 0.7 \pm 0.1$, which taken as face value, implies that the data are consistent with a quickly radially accelerating wind, in agreement with theoretical expectations for stellar feedback-driven wind \citep[see, e.g.,][]{Dong2018}. The wind is initially launched with a velocity $v_{0} = 32 \pm 5$ km s$^{-1}$ and then it increases up to a maximum velocity $v_{\mathrm{max}} = 450 \pm 40$ km s$^{-1}$, which implies a maximum radius reached by the galactic wind of $70 \pm 40$ kpc (through Eq.~\ref{eq:vel_field}), that is consistent with what we see in the MUSE data (see Fig~\ref{fig:mg2nb1}). 

However, a wind velocity that increases with radius does not necessarily imply that the wind is indeed accelerating. If there is a distribution of velocities for the gas leaving the interstellar medium (ISM), then higher velocity material will naturally dominate at larger radii, even if the wind is decelerating at all radii, as shown by the hydrodynamical simulations of SNe feedback-powered outflows from \citet{DallaVecchia2008}, producing a velocity profile that increases with radii. We caution the reader about this possible degeneracy in order to avoid an overinterpretation of our modeling results.

The P-Cygni profile shows a velocity offset $\Delta v = 97 \pm 5$ km s$^{-1}$ with respect to the systemic redshift. This offset can be appreciated in Fig.~\ref{fig:aperspec1}, especially in the innermost apertures, where the \ion{Mg}{II} rest-frame wavelength coincides with the absorption, rather than with the emission peak. On the other hand, in the outer apertures, the \ion{Mg}{II} rest-frame wavelength roughly coincides with the emission line. Thus, this inferred velocity offset is likely a result of the complex kinematic structure of the outflow not being properly described by a single power law. We discuss this further in Sect.~\ref{sec:discussion}.

It is also worth noting the covariance between some of the parameters. For instance, Fig.~\ref{fig:posterior_dist} shows a clear negative covariance between $\gamma$ and $v_{0}$, meaning that for winds that are launched with higher initial velocities, a smaller increase in velocity with radius is needed to reproduce the data. Similarly, $\Delta v$ also shows some covariance with $v_{0}$ and $\gamma$, which is not unexpected since as mentioned above, the velocity offset is closely related to the adopted velocity law (thus, to $\gamma$ and $v_{0}$).

Regarding the nebular contribution to the emission, parameterized by $f_{C}$, we find that a minor contribution is needed to model the data ($\sim12\%$ of the continuum level). However, this might not be entirely correct, since our model is not able to reproduce the observed ratio of the \ion{Mg}{II} emission lines, particularly in the inner apertures, where nebular emission could be present. This limitation is further discussed in Sect.~\ref{sec:discussion}.

\begin{figure}
\includegraphics[width=\columnwidth]{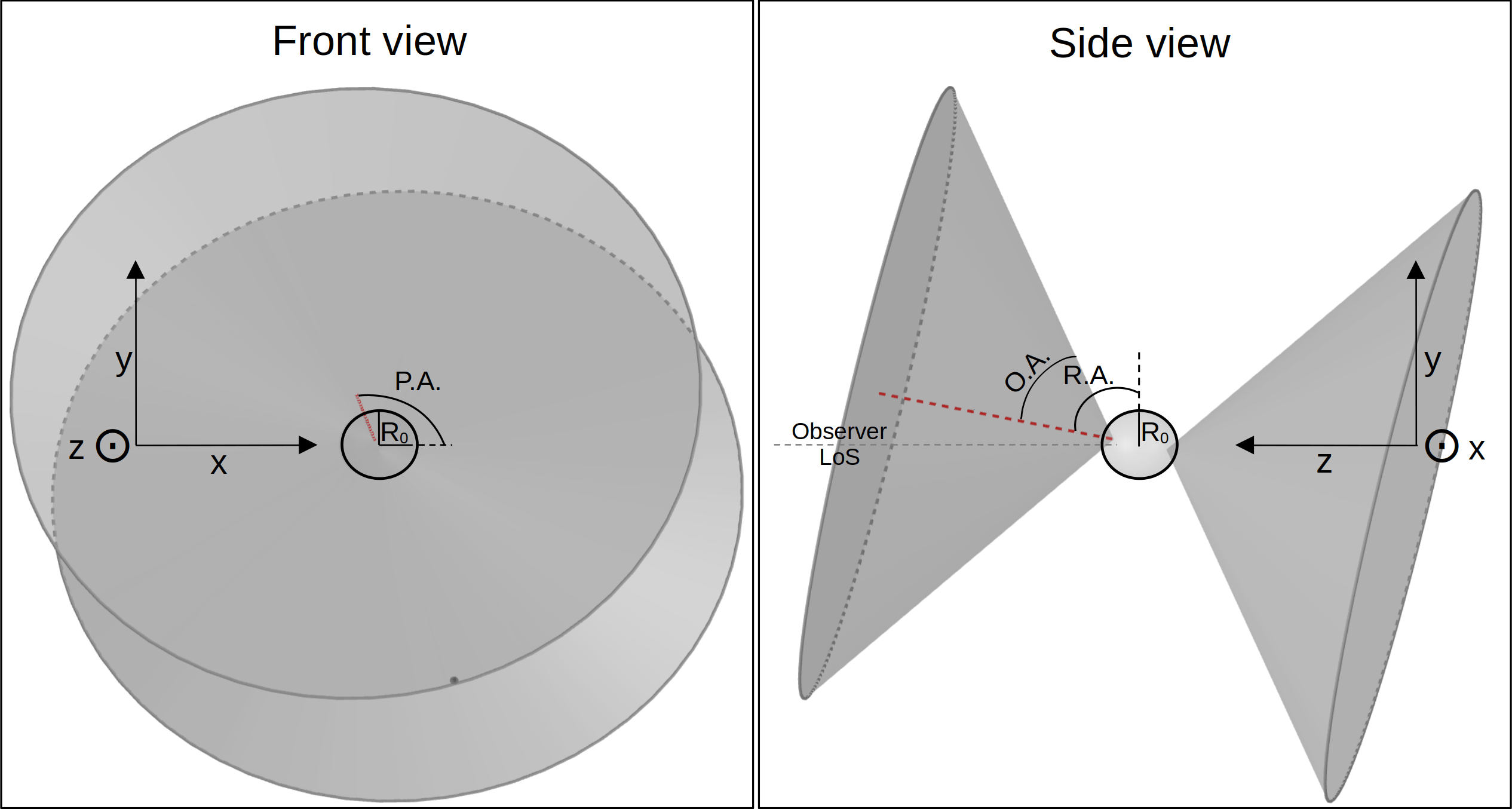}
\caption{Schematic representation of the geometry of our best-fitting outflow model, with O.A. = 59$^{\circ}$, P.A. = 120$^{\circ}$, and R.A. = 80$^{\circ}$, as described in Sect.~\ref{sec:best_fitting_model}. The red dashed line indicates the cone axis. The observer is located at a large positive $z$ value.}
\label{fig:bicone_best_fit}
\end{figure}

Finally, the geometry of the outflow is well constrained by our model. We find a nearly face-on outflow (R.A. = $80\pm10$$^{\circ}$), meaning that we observe the cone almost along its symmetry axis "down the barrel". This coincides with what we would expect from a polar outflow, since \#884 is also seen nearly face-on. Therefore, P.A. = $120\pm10$$^{\circ}$ has little impact on the fitting results. The resulting O.A. of the bicone is large, $59\pm4$$^{\circ}$, and this produces an asymmetric emission line profile that is particularly evident in the outermost apertures in Fig.~\ref{fig:annular_spectra}. In general, the inferred geometry resembles what we see in the data cube, as shown in Fig.~\ref{fig:cont_subs_FoV}. Figure~\ref{fig:bicone_best_fit} provides a schematic representation of the biconical geometry produced by this combination of angles. Table~\ref{tab:best_fit} shows the mean and standard deviation of the posterior distribution of all our model parameters.

\begin{table}
\caption[]{Best-fitting model parameters.}
\begin{tabular}{@{}ll}
\hline\noalign{\smallskip}
Parameter & Value \\
\hline
$R_0$  & $1.6\pm 0.2$ kpc\\
$\tau_0$ & $4000\pm2000$ \\
$\gamma$ & $0.7\pm0.1$ \\
$v_0$ & $32\pm5$ km s$^{-1}$\\
$v_{\mathrm{max}}$ & $450\pm40$ km s$^{-1}$ \\
$\Delta v$ & $97\pm5$ km s$^{-1}$\\
$\log$ $f_{C}$ & $-0.9\pm0.5$ \\
O.A.& $59\pm4$$^{\circ}$ \\
P.A.& $120\pm10$$^{\circ}$ \\
R.A.& $80\pm10$$^{\circ}$ \\
log $f$& $-3.1\pm0.9$ \\

\noalign{\smallskip}\hline\noalign{\smallskip}

\end{tabular}
\label{tab:best_fit}
\end{table}

\subsection{Spectra extracted from annular apertures}

Beyond comparing the \ion{Mg}{II} emission pseudo-narrowband images for the data and model cubes, it is also relevant to compare their respective spectral energy distributions. To do this, we extract and compare the spectra from the same ring-like apertures defined in Sect.~\ref{sec:radial} and shown in Fig.~\ref{fig:mg2nb1}, for the data cube and the best-fitting model cube.
\begin{figure*}
\includegraphics[width=\textwidth]{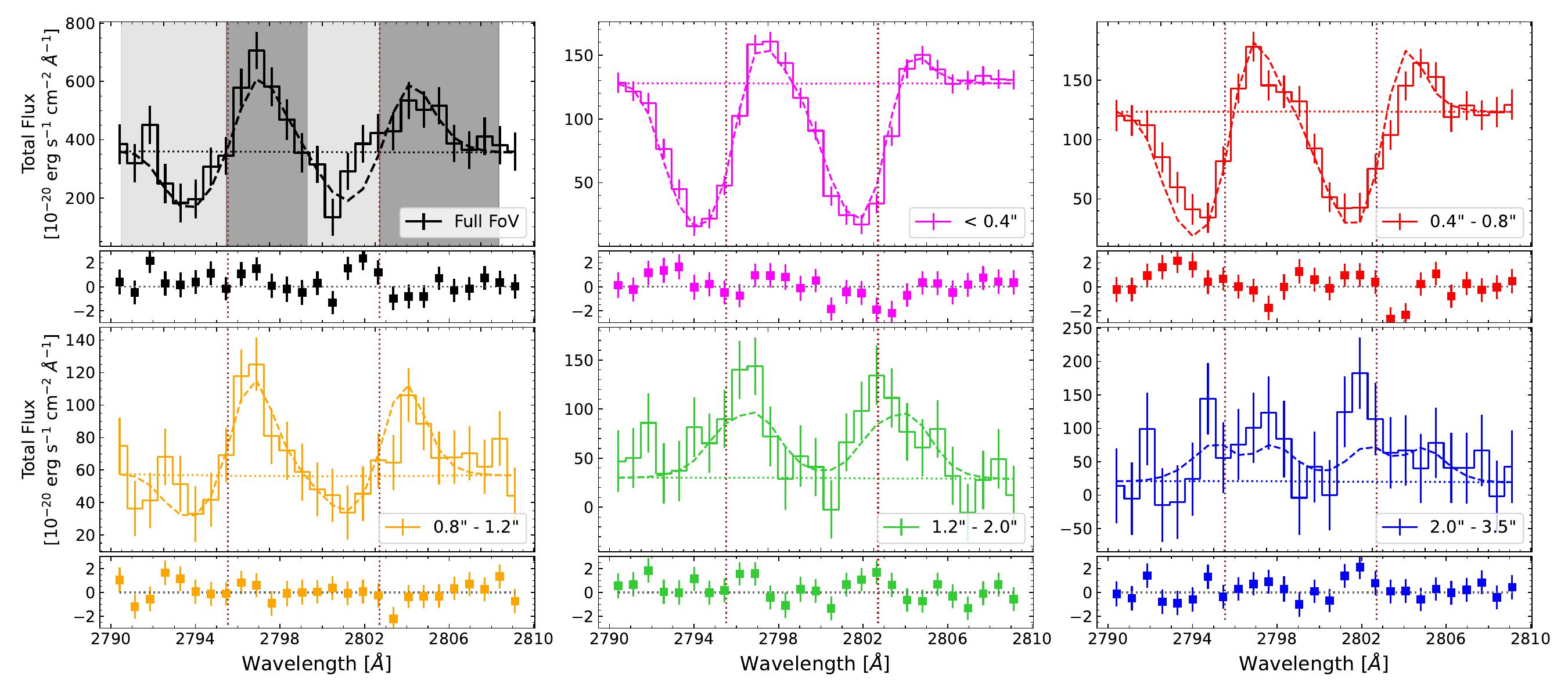}
\caption{Rest-frame spectra extracted from five ring-like apertures as well as from the full modeled FoV for the data cube (solid lines) and the best-fitting model cube (dashed lines). The labels in each panel indicate the inner and outer radii of each ring-like aperture (except the innermost, circular aperture, which is defined by a single radius). In the frame of reference of the central galaxy, the apertures represent physical sizes of $<3$ kpc, $3-6$ kpc, $6-9$ kpc, $9-15$ kpc, and $15-26$ kpc. The nearly horizontal dotted lines show the continuum level for each aperture. The vertical dotted lines indicate the rest-frame wavelength of the \ion{Mg}{II} doublet. The smaller panels show the residuals of the model, normalized by the median standard deviation of the spectra extracted from each aperture. That is, a residual of $1$ indicates one sigma difference between the data and the model. Our outflow modeling scheme reproduces the \ion{Mg}{II} emission and absorption profiles relatively well. The top-left panel shows in gray colors the absorption- and emission-dominated bands (light gray and dark gray, respectively), defined by the wavelengths where the full FoV model intersects the continuum. We use these bands to produce emission- and absorption-narrowband images that we analyze in Sect.~\ref{sec:radial_profiles} }
\label{fig:annular_spectra}
\end{figure*}
 
Figure~\ref{fig:annular_spectra} shows the spectra extracted from these apertures, as well as from the full modeled FoV, for the data (solid lines) and the model (dashed lines) cubes. It is clear that there is relatively good agreement between both cubes, with an inner spectrum dominated by strong absorption features and little emission, and a progressively more emission-dominated spectrum toward the outer apertures. The model successfully reproduces the observed line profiles, especially in the inner apertures. On the other hand, there are some differences between the data and the model that become more obvious in the outer apertures. For instance, the model does not reach the level of emission present in the data cube in the outer apertures. The observed emission lines in the outer regions are slightly but systematically stronger than in the model. Furthermore, the emission in the outer apertures of the data cube is slightly blue-shifted with respect to the inner apertures, and shows in general a more complex kinematic structure. Some of these differences are discussed more in detail in Sect.~\ref{sec:differences_data_model}. Recall that the radial binning of the data and model cubes is only for visualization and comparison purposes. The model is fitted to the data cube, not to aperture-integrated spectra.

\subsection{Radial profiles}
\label{sec:radial_profiles}

In the top-left panel of Fig.~\ref{fig:annular_spectra} we show in gray the absorption- and emission-dominated bands, defined by the wavelengths where the continuum intersects the modeled P-Cyngi profile, for the full FoV spectrum. To compare the spatial distribution of \ion{Mg}{II} emission between the model and the data, we have computed the absorption- and emission-centered narrow-band images, summing the wavelength channels within these absorption- and emission-dominated bands, for the continuum-subtracted data and model cubes. 

\begin{figure*}
\includegraphics[width=\textwidth]{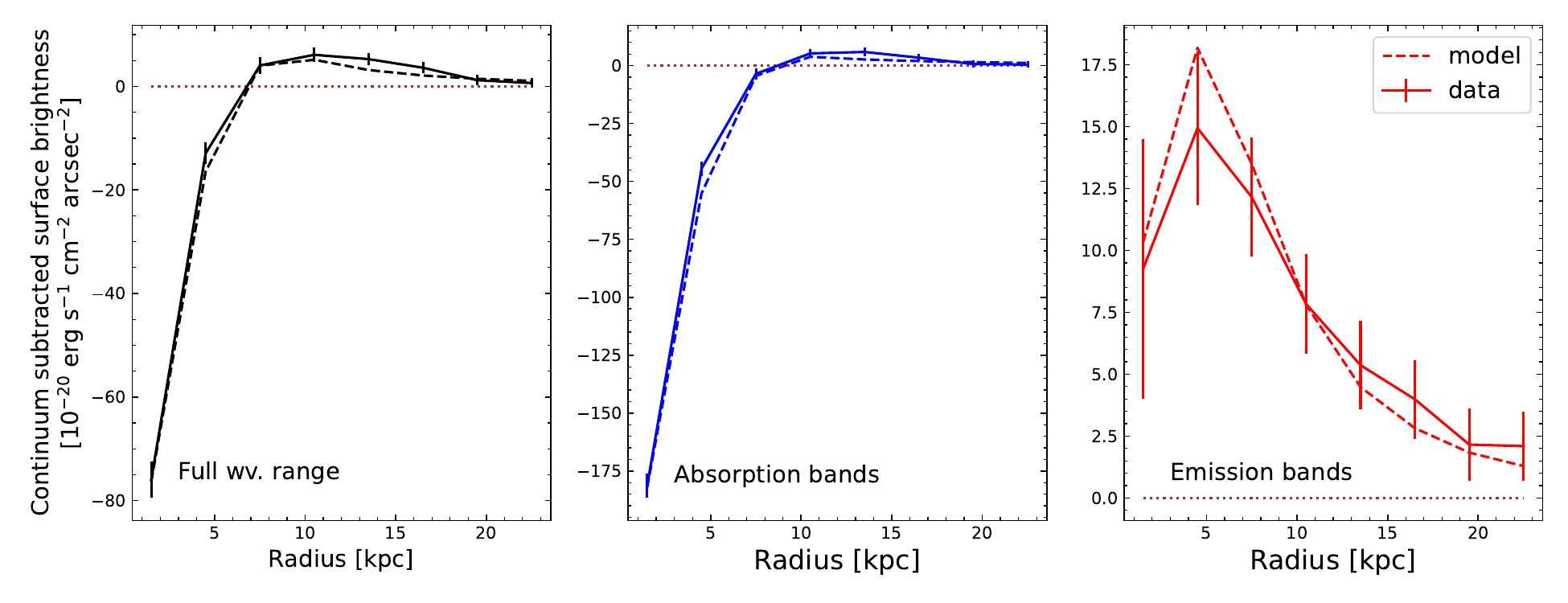}
\caption{Radial profiles of the \ion{Mg}{II} surface brightness measured for the absorption (blue) and emission (red) dominated bands, defined in the top-left panel of Fig.~\ref{fig:annular_spectra}, as well as for the full wavelength range (i.e., total absorption plus emission), for the data cube (solid lines) and best-fitting model cube (dashed lines). The brown dashed line indicates the zero level (i.e., continuum surface brightness). The model reproduces several key features of the radial profiles, such as a similar depth in the absorption in the central region, and a decrease in the emission profile toward outer radii generally consistent within the error bars, with a peak at around $\sim 5$ kpc from the central galaxy.}
\label{fig:radial_profiles}
\end{figure*}

Figure~\ref{fig:radial_profiles} shows the radial profiles of the \ion{Mg}{II} surface brightness measured for the absorption- and emission-dominated bands, as well as for the full wavelength range (i.e., total absorption plus emission), computed for the data and model cubes, using a radial bin size of 3 kpc. The model is able to reproduce various features of the radial profiles, such as a similar depth in the absorption in the central region, and a decrease in the emission profile toward outer radii generally consistent with the data within the error bars. We note that in the left panel, the radial profile in the central region is shallower compared to the middle panel, because the left panel includes the contribution of central emission. On the other hand, the largest difference arises in the inner regions of the emission-dominated narrow-band image, where the model cube exhibits an excess of emission compared to the data cube, specifically at $\sim 5$ kpc radius. This radius matches the second inner aperture (top-right panel) in Fig.~\ref{fig:annular_spectra}. This figure shows that this additional flux comes, at least in part, from the red transition of the \ion{Mg}{II} doublet, since, as mentioned earlier, our model does not reproduce the observed  \ion{Mg}{II}$\lambda 2796$ to \ion{Mg}{II}$\lambda 2802$ ratio. We discuss this discrepancy further in Sect.~\ref{sec:differences_data_model}. 

\begin{figure*}
\includegraphics[width=\textwidth]{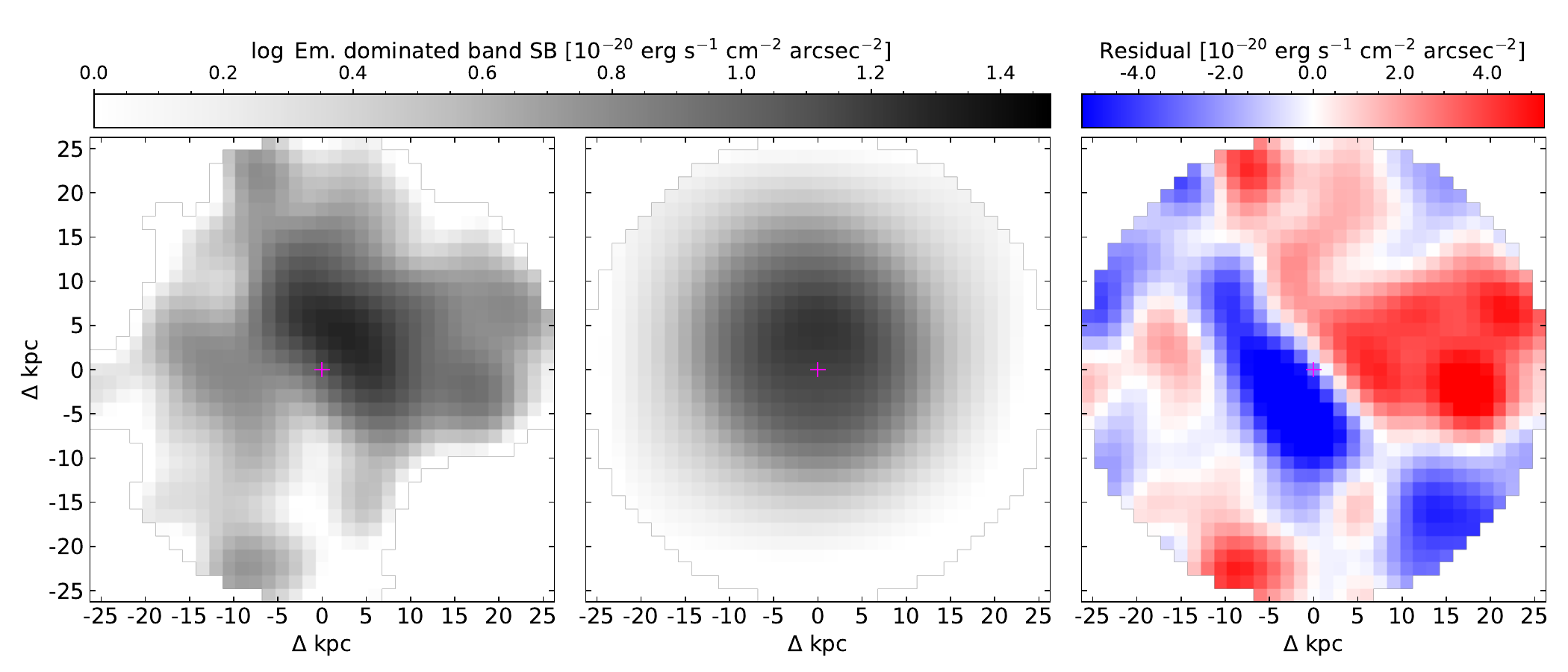}
\caption{Emission-dominated band image of the \ion{Mg}{II} surface brightness, computed from the data (left) and model (center) cubes, convolved with a Gaussian of 1\arcsec\ FWHM. The emission-dominated band is defined in the top-left panel of Fig.~\ref{fig:annular_spectra}. The right panel shows the residual between the two images. The magenta cross indicates the center of the FoV for reference.}
\label{fig:em_band_image}
\end{figure*}

Figure~\ref{fig:em_band_image} shows the emission-dominated band images, used to compute the radial profile in the right panel of Fig.~\ref{fig:radial_profiles}, for the data and model cubes, as well as the corresponding residual. In this band, the data image looks more asymmetric than in Fig.~\ref{fig:cont_subs_FoV}. This is because it shows the inner and brighter region of the \ion{Mg}{II} halo, that in Fig.~\ref{fig:cont_subs_FoV} is dominated by the central absorption. It also should be kept in mind that this figure does not show all the \ion{Mg}{II} emission in the data. A fraction of the emission is blueshifted to the absorption-dominated band, especially in the outer regions (see Fig.~\ref{fig:annular_spectra}). Similarly, in the model image, it is also possible to appreciate better the spatial distribution of the \ion{Mg}{II} emission within the central region, which is overwhelmed by the absorption in Fig.~\ref{fig:cont_subs_FoV}. It exhibits a central dip, and a subsequent radial increase of flux, followed by a radial decrease of \ion{Mg}{II} emission, as shown in the right panel of Fig.~\ref{fig:radial_profiles}. The origin of the central dip is that the physical location (and hence, wavelength) where the strongest peak of the emission should be at such a low impact parameter, is actually a radius $r < R_{0}$, where we assume there is no wind to emit.

\section{Discussion}
\label{sec:discussion}

\subsection{Model limitations}
\label{sec:model_limitations}

\subsubsection{Single scattering approximation}
As explained in Sect.~\ref{sec:basic_model}, if the velocity field is steep enough, the \ion{Mg}{II} photons are only able to interact with the ions in a specific discrete spherical shell, and they can thus easily escape immediately after being reemitted. On the other hand, if the velocity field is closer to flat (i.e., $\gamma \approx 0$), the intrinsic velocity distribution of ions within a given spherical shell (or also the presence of turbulence) would allow the \ion{Mg}{II} photons to interact with ions within a broader layer, thus, making the reabsorption of photons more likely. However, since \ion{Mg}{II} does not present an alternative fluorescent channel, these reabsorbed photons would eventually be reemitted again through the same transition, although possibly at a different wavelength than the originally reemitted photon. This could potentially modify the line shape of the emission component of the P-Cygni profile. 

While implementing a more realistic treatment to model multiple scattering within the outflow is beyond the scope of this paper, as it would require the implementation of full radiative transfer numerical simulations  \citep[see, e.g.,][]{Prochaska:2011eqa}, our results suggest a velocity field with a reasonably well-constrained steepness $\gamma \approx 0.7$. This supports the idea that the single scattering approximation is a justified choice for this case (our approach does not prevent at any point a flatter velocity field, and therefore the steep gradient obtained is not a consequence of our assumptions).

\subsubsection{Complexity of CGM physics}
\label{sec:oversimplification}

A key prediction from cosmological simulations is that the CGM is a dynamic entity, where several different cosmological processes take place  \citep{Voort2012,AnglesAlcazar2017}. There are cold and hot accretion modes of gas from the CGM into the central galaxy (i.e., inflows), satellite galaxies that do not separate cleanly from the CGM of the central galaxy they orbit, and galactic winds driven by stellar or AGN feedback \citep[see the review from][for a comprehensive description of these different phenomena]{Faucher2023}.

The model implemented here consists of an approximation to describe galactic winds, meaning that any contribution of the other physical processes taking place in the CGM of MUSE-HUDF $\#884$ is not described by our model. In the same line, using a single velocity law, parameterized by $\gamma$ across the full extent of the galactic wind is likely unrealistic. A more flexible parametrization of the velocity field could better describe the combination of outflows and inflows present in the CGM. Turbulence is also likely to be present in the CGM \citep{Schmidt2021, Koplitz2023}. Turbulent motion could modify the kinematic structure of the outflowing gas, especially in the outer regions, where gas is less gravitationally bound to the central potential. In the particular case of \#884, it is also possible that tidal forces could play a role in shaping the outer part of the CGM, if the nearby galaxy NB1 is close enough to the \ion{Mg}{II} halo (although the actual distance between both galaxies is unknown, we cannot rule out this possibility)

Furthermore, a combination of observations and simulations suggests that outflowing material from the CGM presents a spatial distribution consistent with clumps of gas varying in strength on physical scales < 2 kpc \citep[see, e.g.,][]{Augustin2021}. Similarly, \citet{Afruni2023} use a sample of \ion{Mg}{II} $\lambda\lambda$2796,2803 absorbers in the MUSE spectra of giant gravitational arcs to infer a coherence scale of the gas in the CGM in the range of $1.4 - 7.8$ kpc, and \citet{Dylan2020} report similar conclusions using purely cosmological magnetohydrodynamical simulations. This contrasts with the smoothly varying gas density distribution assumed in our model. Hence, reproducing small-scale features of the CGM is beyond the scope of our model, although this becomes less important when we compare spatially averaged quantities (e.g., aperture-integrated spectra), such as in Sect.~\ref{sec:radial_profiles}.

Additionally, due to this smooth underlying gas distribution, we implicitly assume a partial covering fraction for the absorption of $1$ within the biconical outflow, which contrasts with the radially varying covering fraction reported by \citet{Martin:2009fl} and \citet{Steidel2010}. However, our model accurately reproduces the spectral profile of the absorption in the inner-most aperture, and thus, we do not find strong evidence that suggests a lower covering fraction (although, due to the LSF, there could be some degree of degeneracy between a lower optical depth and a covering fraction lower than 1).

\subsection{Comparing data and model cubes}
\label{sec:differences_data_model}
While there is a general agreement between the best-fitting model and the data cubes, when it comes to comparing their radially resolved spectra and radial profiles, the relative simplicity of the model leads to some higher-order differences between them that should be subject to closer inspection:

The model does not accurately reproduce the inner radial profile of the emission-dominated band image (see the right panel of Fig.~\ref{fig:radial_profiles}). The model exhibits excess emission that peaks at about $\sim 5$ kpc from the central source.  Although this difference could, in principle, be caused by noise, since the inner radial bins are more strongly affected by noise due to the lower number statistics (i.e., fewer pixels), it is more likely related to the model not reproducing the observed relative strength of the two \ion{Mg}{II} transitions, putting a relative excess of flux in the red line of the doublet as a result (see also the top right panel of Fig.~\ref{fig:annular_spectra}). Although by construction, we use half of the optical depth $\tau_{0}$ for the \ion{Mg}{II} $\lambda$2803 line, this hardly makes any difference at such high overall optical depth (due to the factor $e^{-\tau}$). Thus, a stronger \ion{Mg}{II} $\lambda$2796 emission in the data (compared to the \ion{Mg}{II} $\lambda$2803 line) could hint at unaccounted-for nebular emission. The fact that in the outer apertures, where the continuum contribution is virtually zero, the ratio between the two lines is closer to one provides further evidence that this is the case. This would be in agreement with the findings from \citet{Seon2023} for an expanding spherical outflow, where the authors use radiative transfer simulations to explore radial variations in the \ion{Mg}{II} $\lambda$2796-to-$\lambda$2803 ratio induced by resonant scattering. As for why this contribution is not fully captured by our model, an explanation could be that it is of second order, compared to the continuum scattered emission, and harder to constrain given our flux uncertainties, although our best-fitting model already accounts for some nebular contribution. In addition, differences in orientation and occultation of light on disk scales could also contribute to enhancing discrepancies between the data and the model in the inner part of the radial profiles.

The emission in the outer apertures of the data cube is blueshifted with respect to the central regions, and the model is not able to reproduce this feature, as it would require a more complex description of the gas kinematics than a simple power law (see Sec~\ref{sec:oversimplification}). This discrepancy is evident in the spectra extracted from the two outer apertures in Fig.~\ref{fig:annular_spectra}, where the observed emission is systematically bluer than the model emission in the same apertures, and also bluer than the observed emission in the inner apertures. A similar feature was reported for circumgalactic Ly$\alpha$ emission by \citet{Guo2023}, where the authors attributed it to the inner part of the haloes being dominated by resonant scattered Ly$\alpha$ photons from the outflowing gas, while inflows gradually become more relevant toward the outskirts of the haloes (although they do not rule out the contribution of satellites). \citet{Turner2017} and \citet{Chung2019} compared a sample of Lyman Break Galaxies to hydrodynamical simulations, and also concluded that cold accretion flows dominate the CGM at larger impact parameters.

The data cube presents stronger emission in the outer parts ($r > 9$ kpc), compared to the model cube. A discrepancy in the emission toward larger radii is not unexpected, since an incomplete description of the gas kinematics (as discussed above) also yields differences in total flux. This is because the gas density (and, thus, optical depth) depends primarily on the velocity field assumed (through Eq.~\ref{eq:density}). 
    
The observed \ion{Mg}{II} halo does not present a smooth spatial distribution. This is evident in Fig.~\ref{fig:mg2nb1} for the outer parts of the halo, and in Fig.~\ref{fig:em_band_image} for the inner parts. Instead, it exhibits a more irregular distribution, consistent with previous findings from observations and simulations (see discussion in Sect.~\ref{sec:oversimplification}). However, our model, as described in Sect.~\ref{sec:model}, is not able to reproduce such a configuration due to the underlying spherical symmetry assumed in Eq.~\ref{eq:density}.

Nevertheless, what we learn from the differences between the data and our simple but physically motivated model is as relevant as what we learn from their similarities. Our model shows that outflows are most likely biconical, but that the CGM cannot be described by a simple biconical outflow, as suggested by the residuals. Therefore, alternative mechanisms must play a role in shaping the CGM gas, as discussed in previous sections.

\subsection{Strong \ion{Mg}{II} absorption at 30~kpc from the galaxy}
\label{sec:bg1abs}

As mentioned in Sect.~\ref{sec:discenv}, the spectrum of the background galaxy BG1 ($z = 1.221$, $m_\mathrm{AB, F775W} = 24.5$, projected impact parameter for \#884 of 30~kpc) shows strong \ion{Mg}{II} doublet absorption at the same redshift as \#884. 

These absorption features are likely produced by outflowing or inflowing material in the circumgalactic medium of \#884. A third possibility is that the absorption is tracing quasi-static gas, that is, no significant radial movement, as reported by \citet{Weng2023}. We use the spectra of BG1 to compare the absorption predicted by our best-fitting model at that location with the observed absorption.

This comes with the caveat that the galaxy BG1 is beyond the modeled radius, so this implies an extrapolation of the model. In previous sections we have established a number of differences between the data and model, that are mostly relevant in the outer apertures. Hence, BG1 provides an independent sight line to test our model and find out what the discrepancies are between the data and the model toward even larger impact parameters, where the emission of \ion{Mg}{II} is more difficult to probe due to contamination from nearby galaxies (such as BG1) and the overall lower S/N. While our model was adjusted to fit the data cube at impact parameters $<26$ kpc, it can, in principle, be used to infer the distribution of the outflowing gas toward larger distances, and evaluate if a simple outflow model is still a valid approximation to describe the outer CGM.

\begin{figure}
\includegraphics[width=\columnwidth]{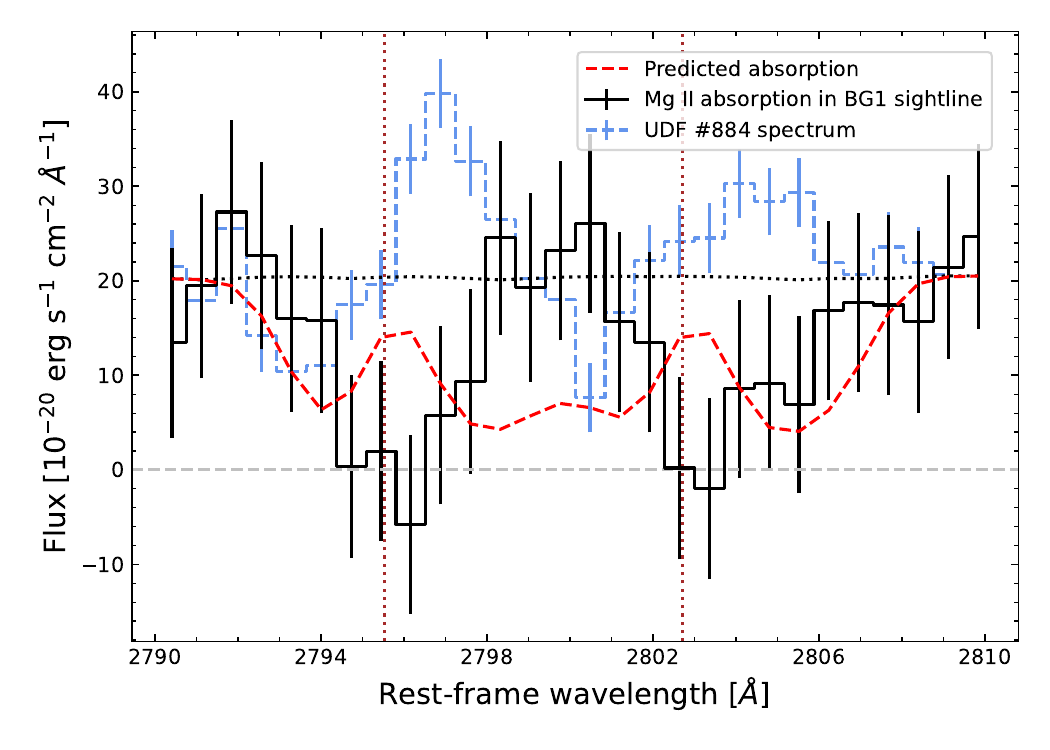}
\caption{Doublet absorption of \ion{Mg}{II} at the redshift of \#884 in the spectrum of the galaxy BG1 (black). The horizontal dotted line shows the continuum level in the spectrum of BG1. The vertical dotted lines show the rest-frame wavelengths of the \ion{Mg}{II} doublet. The red dashed line shows the absorption line predicted by extrapolating our best-fitting model to the location of the BG1 sight line. The blue dashed line shows the full FoV spectrum of \#884, with the continuum level renormalized to match that of BG1. The absorption in the spectrum of BG1 is considerably broader than that present in the spectrum of \#884. This is because in the case of BG1, both sides of the spherical outflow absorb flux, instead of only the blue-shifted half of the outflow. The predicted absorption is inconsistent with the observed profiles, implying that our outflow model does not describe the gas distribution properly along this sight line.}
\label{fig:BG1_spec}
\end{figure}

Thus, we evaluate our model at the BG1 sight line to check if the observed absorption strength is consistent with our predicted value. Figure~\ref{fig:BG1_spec} shows the \ion{Mg}{II} doublet absorption at the redshift of \#884 in the spectrum of the galaxy BG1. The absorption lines are nearly completely saturated, and slightly redshifted compared to the absorption present in the \#884 spectrum, by about $65$ km s$^{-1}$. The absorption in the spectrum of BG1 is also considerably broader than that present in the spectrum of \#884. This is because in the case of BG1, both sides of the biconical outflow absorb flux, instead of only the blue-shifted half of the outflow, as for \#884. The figure shows that although the absorption strength predicted by our model (red) is not negligible (more than $\sim70\%$ of the continuum), it is significantly weaker than the observed absorption. In Sect.~\ref{fig:annular_spectra}, we show that our model underestimates the total emission (hence, gas density) in the outer apertures. Therefore, it is not surprising that it also underestimates the absorption strength toward BG1 at an impact parameter of $\sim30$ kpc. 

Additionally, the modeled absorption is considerably broader than the observed. This hints possibly at an overestimation of the maximum radius by the model. Indeed, a maximum radius of $40$ kpc (which would correspond to a $V_{\mathrm{max}}$ of $\sim330$ km s$^{-1}$, keeping other parameters unchanged) produces absorption features that match the width of the absorption in the spectrum of BG1 much better. This value is, however, significantly lower than the $V_{\mathrm{max}}$ derived from the inner regions of the outflow (although $40$ kpc is indeed consistent within the error bars with our inferred $R_\mathrm{max} = 70\pm40$ kpc), suggesting again that our simple prescription to model the velocity of the gas is not able to describe the more complex kinematics and/or morphology of the outer regions of the CGM of \#884.

Finally, the biconical geometry of the modeled outflow, that produces the asymmetric line profiles, does not provide an accurate description of the absorption features at this large impact parameter. Qualitatively, the absorption features are closer to symmetric, hinting to a spherical configuration of the CGM in the outer region (or at least, a more continuous distribution of gas along this discrete sight line). An alternative could be that this absorption is probing a cold and spatially coherent inflow of gas \cite[this idea would also be consistent with the fact that BG1 is located nearly perpendicular to the major axis of the outflow in Fig.~\ref{fig:mg2nb1}; see, e.g.,][]{Weng2023}

Altogether, this comparison confirms that while our outflow model performs well in reproducing the spectral and spatial properties of the inner ($< 26$ kpc) regions of the CGM \ion{Mg}{II} emission, it lacks the complexity necessary to provide an accurate representation across the full extent of the CGM, where other mechanisms (e.g., inflows) can be present, or even dominant. 


\subsection{Mass outflow rate and mass loading factor}
\label{sec:outflow_mass_rate}

Previous works have aimed at inferring the amount of outflowing gas by estimating the radius at which the optical depth becomes $\sim 1$, and the velocity gradient of the outflow from the width of the observed spectral lines \citep[e.g.,][]{Martin:2013ho, Guo2023b}. In this analysis, our outflow model provides a self-consistent mass outflow rate that results from the combination of the velocity law and density/optical depth profile described in Sect.~\ref{sec:basic_model}. Thus, at any point in the 3-D space, we can infer the density and the velocity of the gas from our analytic framework. Here, we present and discuss the implications of the constraints on the mass outflow rate inferred from our outflow model.

The density of the outflowing gas is directly proportional to the inferred optical depth through Eq.~\ref{eq:tau_0_n_0}. Replacing the constants, we have
\begin{equation}
    \tau_0 = 4.57\times10^{-7} \mathrm{cm}^{3} \mathrm{s}^{-1} n_0 \frac{R_0}{v_0},
\end{equation}
which leads to $n_0 = 5.76\times 10^{-6}$ cm$^{-3}$, using the best-fitting model values of $\tau_0$, $R_0$, and $v_0$. However, in this context, $n_0$ corresponds to the number density of ionized Mg ($n_{\mathrm{Mg^{+}},0}$) at radius $R_{0}$. To compute the corresponding hydrogen density of the gas ($n_{\mathrm{H},0}$), it is necessary to correct by metallicity $\eta(\mathrm{Mg})$, dust depletion $d(\mathrm{Mg})$ and ionization fraction $\chi(\mathrm{Mg^{+}})$ \citep[see, e.g., ][]{Martin:2013ho}.  Specifically, $n_{\mathrm{H},0}$ is given  by
\begin{equation}
\label{eq:density_correction}
    n_{\mathrm{H},0} = \frac{n_{\mathrm{Mg^{+}},0}}{\eta(\mathrm{Mg})\,\chi(\mathrm{Mg^{+}})\,d(\mathrm{Mg}) }.
\end{equation}

Assuming solar metallicity leads to $\eta(\mathrm{Mg}) = 3.8\times10^{-5}$. While the emission line ratios of \#884 would lead us to expect slightly subsolar abundances, we stick for simplicity to the solar value since the outflowing gas might be more enriched with metals. We discuss the uncertainties in these assumptions below.

For dust depletion, we choose as our fiducial value the depletion factor reported by \citet{Guseva:2019hv}, measuring the Mg-to-Ne abundance ratios for a sample of $\sim4200$ galaxies from the SDSS DR 14 survey \citep{SDSS14}, which for a galaxy of the mass of \#884, would correspond to $\approx 0.3$ dex.

Finally, the ionization fraction correction is a function of the ionization parameter $U$. We estimate a value of log $U$ $\simeq$ -3.25, anticipating the evaluation of several optical nebular emission line ratios ([\ion{Ne}{III}]/[\ion{O}{II}], [\ion{O}{III}]/[\ion{O}{II}], and [\ion{O}{III}]/H$\beta$) in the follow-up study of this galaxy by Wisotzki et al. (in prep.). 

We used then the relation between $\chi(\mathrm{Mg^{+}})$ and $U$ reported by \citet{Martin:2013ho}, determined using Cloudy \citep{Ferland2013}, to compute the corresponding ionization fraction:
\begin{equation}
    \chi(\mathrm{Mg^{+}}) = 4 \times 10^{-4} U^{-0.94},
\end{equation}
which leads to an ionization fraction correction $\chi(\mathrm{Mg^{+}}) \approx 0.45$. Correcting for metallicity, dust depletion and ionization fraction using this set of assumptions allows us to calculate a total hydrogen number density of $n_{\mathrm{H},0} = 0.67$ cm$^{-3}$, following Eq.~\ref{eq:density_correction}.



The outflow mass rate at a given radius $\dot M(r)$ can be expressed as the product between the gas density, the surface area of the outflow ($A$), and the velocity of the outflow:
\begin{equation}
\dot M(r) = \mu\,m_{\mathrm{H}}\,n_{\mathrm{H}}(r)\,A(r)\,v(r),
\end{equation}
with $\mu\,m_{\mathrm{H}}$ equal to the average mass per hydrogen atom. We adopt a value of $\mu = 1.4$, which is commonly used in this context \citep[see, e.g.,][]{Rupke2005, Weiner:2009cf, Martin:2012dx, Burchett2021}. Since we have imposed mass conservation, the expression above does not depend on the radius and can be written as
\begin{equation}
\label{eq:outflow_rate}
    \dot M = 4\,\pi\,\mu\,m_{\mathrm{H}}\,n_{\mathrm{H},0}\,C_{\Omega}\,R_{0}^{2}\,v_{0},
\end{equation}
where $C_{\Omega} = 1 - cos(\mathrm{O.A.})$ corresponds to the angular covering fraction of the bicone. Evaluating the equation yields a mass outflow rate of
\begin{equation}
\label{eq:mass_loading}
\dot M \approx 12\pm7\,\mathrm{M}_{\odot} \,\mathrm{yr}^{-1},
\end{equation}
which implies a mass loading factor $\dot M / \mathrm{SFR} \approx 1.0$ (considering the uncertainties in the SFR). Equation~\ref{eq:mass_loading} shows only the propagated statistical uncertainty, where the main source of uncertainty in $\dot M$ comes from the uncertainty in $\tau_{0}$, and hence, $n_{\mathrm{H},0}$. However, besides the uncertainties propagated from the determination of our best-fitting parameters, the mass outflow rate and mass loading factor are also subject to relevant systematic uncertainties that result from the assumptions made to calculate the total hydrogen number density from the measured optical depth of ionized magnesium.

For instance, while we have assumed a solar metallicity to perform our corrections, the mass vs. gas-phase metallicity relation reported by \citet{Zahid2014} suggests a metallicity of $\sim2\,Z_{\odot}$ for the ISM of \#884, with an average scatter of $0.3$ dex \citep{Tremonti2004}. Furthermore, \citet{Chisholm2018} find that the metallicity of outflowing gas is metal-enriched with respect to the ISM of star-forming galaxies, by a factor $\sim3$ for galaxies with mass similar to that of \#884. On top of this potential systematic difference, it must also be taken into account that the gas-phase metallicity measurements strongly depend on how the metallicity is calculated \citep{Kewley2008}. Absolute metallicity calibrations can vary by up to $0.7$ dex for different calibration systems (and thus, the intercept of the mass vs. gas-phase metallicity relation).

Similarly, there are different estimations for the Mg depletion in the literature. While here we assume the value determined by \citet{Guseva:2019hv}, \citet{Savage1996} find that typical Milky Way clouds have much lower $d(\mathrm{Mg}) =  6.3 \times 10^{-2}$. This would introduce a systematic difference of a factor $\sim8$ in the inferred hydrogen density, that would linearly increase the inferred outflow mass rate. Nevertheless, given that the value reported by \citet{Guseva:2019hv} is based on a large sample of galaxies that encompasses the redshift and mass of \#884, we find that their results are more accurate for our case, and thus, that the extreme case of a factor $\sim8$ with the real underlying Mg depletion is very unlikely.

On the other hand, since we are able to directly constrain the ionization parameter $U$ through the emission lines measured in the spectrum of \#884, we find it unlikely that the real underlying ionization correction is significantly different from our estimated $\chi(\mathrm{Mg^{+}})$. Furthermore, our estimation is also consistent with the range of values calculated by \citet{Murray2007} and \citet{Guseva:2013jp} (although the former derived this quantity for a sample of ULIRGs). However, it is also possible that the \ion{Mg}{II} is shielded from the ionizing radiation, leading to a lower underlying fraction of Mg$^{+}$ ions.

Overall, the measured outflow rate and mass loading factor could be up to a factor $\sim6$ lower, if we assume a gas-phase metallicity given by the relation reported by \citet{Zahid2014}, and boosted according to the findings from \citet{Chisholm2018} (with the caveat that a different metallicity calibration could potentially shift the relation from \citet{Zahid2014}). On the contrary, it could also be up to a factor $\sim8$ higher, using the dust depletion reported by \citet{Savage1996}. Despite the fact that we do not account for additional systematic variations induced by our ionization correction choice, the metallicity and dust depletion choices come with significant systematic uncertainties that must be taken into account in order to properly interpret our results and compare them with the results from different works. 

In the following, we proceed similarly to \citet{Burchett2021} to investigate whether this outflowing material is able to escape the galactic potential or will fall again into the galaxy. We use the stellar-to-halo mass relation from \citet{Moster2013} to calculate a total halo mass of $\log M_{\mathrm{h}} / \mathrm{M}_{\odot} = 11.9$. For a halo of this mass with a Navarro–Frenk–White profile, and assuming a concentration parameter given by the relation from \citet{Ragagnin2019}, we calculate an escape velocity of $\sim 250$ km s$^{-1}$ at the maximum radius reached by the outflow of $\sim70$ kpc. This is significantly lower than the terminal velocity of the outflow $v_{\mathrm{max}} = 450$ km s$^{-1}$, meaning that the outflowing material will likely escape the gravitational potential of the halo and enrich the intergalactic medium, rather than falling again into \#884. However, as shown in previous sections, a radius of $70$ kpc is well beyond the radius where our outflow model describes the data faithfully. If we compare the escape velocity with the outflow velocity at a smaller radius, where our outflow model still performs well, we find that, for instance, at $25$ kpc, the escape velocity is still higher (although comparable, within the error bars) than the outflow velocity ($\sim250$ and $\sim220\pm70$ km s$^{-1}$, respectively). Specifically, given our set of best-fitting parameters, we find that the outflow starts exceeding the escape velocity of the galactic potential at $\sim30$ kpc. Thus, whether the outflowing gas escapes the galactic potential or not depends on the radii up to which our outflow model provides an accurate representation of the data. Our analysis suggests that this holds for impact parameters of $\sim 20 - 25$ kpc. However, this cannot be translated directly into radii due to the additional line-of-sight component of the distance, and the face-on nature of the outflow. Moreover, we find evidence of high-velocity gas ($\sim400$ km s$^{-1}$) in the down-the-barrel spectrum of the outflow. Altogether, we find it more likely that our outflow model still performs well by the point where the outflow velocity overcomes the escape velocity of the galaxy halo and, thus, that the outflowing gas will be able to escape the galaxy.

Finally, solving the differential equation given by Eq.~\ref{eq:vel_field}, it is possible to study the timescales of the outflowing gas, and we find that any given parcel of gas ejected at velocity $v_{0}$ from a radius $R_{0}$ will take $\sim350$ Myr to reach a velocity $v_{\mathrm{max}}$, at a radius $R_{\mathrm{max}}$, considering our best-fitting value of $\gamma = 0.7$. This timescale is significantly longer than the typical star formation timescales of only tens of Myr \citep{Chevance2020}, meaning that there could be large variations in the amount of outflowing material across the outflow due to changes in the star formation rate (i.e., the mass conservation assumption might not hold). Indeed, \citet{Mehta2023} find that galaxies in the mass range of \#884 exhibit predominantly a bursty star formation rate, rather than a smooth one. Nevertheless, although the mass outflow rate inferred from the initial conditions of the outflow $v_0$, $R_0$, and $n_0$ (as in Eq.~\ref{eq:outflow_rate}) might not actually reflect the real amount of gas currently being expelled at $R_\mathrm{max}$, it provides a reasonable reference value consistent with the observations, in the framework of our outflow modeling scheme. 

Our measured mass loading factor is generally within the range of values previously reported in the literature \citep[see, e.g.,][]{Heckman2015, Chisholm2017, Burchett2021, KadoFong2024, Vasan2024}. \citet{Guo2023b} find a mass loading factor of $\sim7$, but they use the dust depletion correction from \citet{Savage1996}. If this difference is accounted for, the values are perfectly consistent, which highlights how relevant is having these different assumptions in mind when contrasting results from different works. Compared to higher redshift galaxies, our results are consistent with the measurements from \citet{Davies2019} and \citet{Freeman2019}, but somewhat higher than the value of $\sim0.05$ reported by \citet{Concas2022}. Our results are also within the range of values predicted by simulations \citep[][]{Dylan2019b, AnglesAlcazar2017, Mitchell2020}.

On the other hand, there are some studies that report significantly larger mass loading factors \citep[see, e.g., ][]{Martin:2013ho, Shaban2023, Lin2023}. However, performing a direct comparison is not straightforward, not only because of the different assumptions made to compute these values, but also due to the different nature of the approaches employed. Further work is required to understand the origin of the dispersion in the measurements of the outflow mass rates.


\subsection{Modeling of the CGM in the literature}
In this section, we place our model results in the context of what previous works have found by comparing simulations and models to CGM observations.
\\
\subsubsection{Comparison to analytical models}
\citet{Fielding2017} use three-dimensional hydrodynamical simulations to study the launching of galactic winds by supernovae and find launching velocity in the order of 10s of km s$^{-1}$, and terminal velocities that range between 300 and 500 km s$^{-1}$, which is consistent with our results.
On the other hand, \citet{Afruni2021}  use semi-analytical parametric models to describe the cool CGM as an outflow of gas clouds from the central galaxy, and concludes that a supernova-driven outflow model is not suitable to describe the dynamics of the cool circumgalactic gas. However, this result is not necessarily in conflict with ours, as the sight lines used by them to compare the model to observations are located at significantly larger impact parameters ($\sim100$ kpc) than those considered in our modeling. Due to simple geometrical effects, sight lines that probe the CGM in absorption tend to cover preferentially larger impact parameters, and it is well known that the CGM is the scenario of several different physical processes \citep[see, e.g.,][]{Faucher2023}, and that these processes will be more or less relevant in different regions of the CGM \citep[e.g.,][]{AnglesAlcazar2017}. In this context, in a recent paper, \citet{Weng2023}  use the TNG50 cosmological, magnetohydrodynamical simulation from the IllustrisTNG project \citep{Dylan2019} to trace back the origin of the CGM particles, finding, for instance, that the contribution of satellite galaxies becomes relevant (or dominant) beyond $\sim60$ kpc (depending on the mass of the central galaxy), and that the ratio of outflowing-to-inflowing gas also depends on the azimuthal angle with respect to the central galaxy, with outflows being more predominant along the semi-minor axis of galaxies.
\\
Along the same line, \citet{Chung2019} compare Ly$\alpha$ observations of the CGM of Lyman Break Galaxies with simulated CGM observations, and find that their simulation, which includes strong, momentum-driven galactic outflows, better reproduces the observed characteristics of the CGM compared to a less efficient feedback model. This is in line with our results, since we are able to reproduce several features of the data with our outflow modeling scheme. However, the simulation still struggles to replicate the observed data of the inner CGM, suggesting other mechanisms may be involved. Their results also indicate that galactic outflows primarily affect Ly$\alpha$ absorption and emission around galaxies within 50 kpc, while cold accretion flows dominate at larger distances. 
\\
\subsubsection{Comparison to results from studying the CGM in absorption}
Absorption lines have been the primary way to study the CGM during the last decade (e.g., results from the COS-Haloes \citep{Werk2013} or MEGAFLOW \citep{Schroetter:2016bl} surveys). In particular, the MEGAFLOW survey has provided a crucial understanding of the non-isotropic distribution of the CGM. \citet{Schroetter:2019es} use a sample of 79 strong \ion{Mg}{II} absorbers, and find that cool halo gas exhibits a strong bimodal distribution in the azimuthal angle of the background QSO sight line, with respect to the galaxy major axis, which suggests a biconical outflow scenario that coexist with a coplanar gaseous structure. These results are further confirmed by \citet{Zabl:2019ija}. Similarly, \citet{Zabl:2020js} use two QSO sight lines to study the CGM of a star-forming galaxy at $z\approx0.7$, and find that the data are consistent with an anisotropic ejection of gas from the galaxy, with a biconical outflow perpendicular to the galaxy disk.

A number of other works have modeled CGM \ion{Mg}{II} absorption as biconical outflows \citep[e.g.,][]{Nielsen2020, FernandezFigueroa2022}. Our results in terms of geometry and velocity of the gas are indeed comparable to those obtained by \citet{Nielsen2020}. However, a one-to-one comparison of our results with others from the literature is difficult, since absorption-based analyses have some intrinsic limitations, compared to a fully spatially resolved modeling of the CGM in emission.

\subsubsection{Comparison to results from studying the CGM in emission}
\citet{Zabl2021} report the first detection of extended \ion{Mg}{II} emission in the halo of a galaxy that is also probed in absorption by a QSO sight line. Their analysis suggests a biconical outflow with an opening angle of $35^{\circ}$ and an outflow velocity of 130 km s$^{-1}$. While this outflow velocity is almost a factor of three higher than our measured $v_0$, there is a fundamental difference in the assumptions made to obtain these numbers; here, we assume a radially accelerating wind, whereas in \citet{Zabl2021}, the authors \citep[motivated by the results from][]{Bouche:2012ho} compute this value by matching a constant velocity outflow model to the data, aiming to reproduce the absorption and emission features observed (although they also explore the possibility of radially accelerating and decelerating winds, and find that the observed surface brightness of the halo can be reproduced with either of these alternative models). Under this assumption, an outflow velocity significantly higher than our wind launching velocity is expected. Furthermore, the constant wind velocities measured by \citet{Bouche:2012ho} from a sample of 11 \ion{Mg}{II} absorbers are not necessarily in conflict with our results, given that the impact parameters of the QSO sight lines to the central galaxies used to obtain this result are significantly larger than the spatial extent of our outflow model. \citet{Zabl2021} also find that a \ion{Mg}{II} halo emission powered only by scattering of continuum photons leads to an inconsistency in the density inferred from the emission and from the absorption (at an impact parameter of $\sim40$ kpc). This result is qualitatively similar to the discrepancy we find between the observed and predicted absorption in the spectrum of BG1 (see Sect.~\ref{sec:bg1abs}).

 \citet{Burchett2021} present another study of the CGM in emission. They investigate a single galaxy at $z = 0.694$, similar to \#884, observed with the KCWI, that exhibits spatially resolved \ion{Mg}{II} CGM emission extending up to a full size of $\sim37$ kpc, with a clear P-Cygni profile in the regions where stellar continuum is present. While this work is similar to what we present in this paper (albeit with a somewhat lower spatial resolution), their conclusions differ from ours. For instance, they compare their data to a suite of 3D radiative transfer models of galactic outflows, and find that their observations are more consistent with an isotropic outflow, and radially decelerating wind. However, due to the computational expense of the radiative transfer simulations, they only explore a very limited region of the full parameter space, in terms of velocity and geometry of the wind. 
In addition, while in our analysis we do not explore negative values of $\gamma$ (i.e., a radially decreasing wind velocity), we find that the width of the model emission lines in the outer regions is consistent with the observed emission lines. On the other hand, \citet{Burchett2021}, find that a radially increasing velocity wind would produce emission lines that are too broad compared to their observations in the outer regions; hence, a decelerating wind is required. Whether this is a particularity of the galaxy they model, or a result of exploring a limited region of the parameter space, in terms of geometry and wind velocity, has yet to be determined.

Employing a different approach, \citet{Erb2012} use near-UV spectroscopy to measure the velocity offset of \ion{Mg}{II} and \ion{Fe}{II}, with respect to the systemic velocity, to investigate large-scale outflows in a sample of star-forming galaxies. They find an average maximum outflow velocity of $\sim730$ km s$^{-1}$, which is somewhat higher than our measured value, with higher velocities observed in high-mass galaxies. For the low mass galaxies ($M_{\star} < 1.1 \times 10^{10}$M$_{\odot}$), they measure a maximum velocity of $\sim470$ km s$^{-1}$, consistent with our derived $v_{\mathrm{max}}$ (although the stellar mass of \#884 is somewhat higher than this threshold). The dependence of the outflow velocity on the galaxy mass is in agreement with predictions from simulations \citep[see, e.g.,][]{Mitchell2020}. \citet{Dylan2019b} use the TNG50 simulation to investigate galactic outflows driven by supernovae as well as supermassive black hole feedback. They also find that outflows exhibit strong collimation and align preferentially with the minor axes of galaxies. The outflow velocity increases with stellar mass and with redshift. Furthermore, galaxies above the star-forming main sequence drive faster outflows, although this correlation inverts at high mass with the onset of quenching, where accreting black holes drive the strongest outflows.

This is encouraging to expand our modeling to a larger sample of galaxies, covering the parameter space in terms of stellar masses, star formation rates, and redshift, and explore the existence of trends between galaxy properties and CGM properties, from an observational point of view. In a future paper, we plan to extend our sample, and provide the first spatially resolved modeling of the CGM in emission, for a statistical sample of galaxies.

\subsubsection{Comparison with results from stacked integral field data}
An alternative approach to study the CGM emission traced by \ion{Mg}{II} in a spatially resolved fashion consists of stacking the extended emission of a large number of galaxies, instead of focusing on an individual source. This effectively increases the signal-to-noise of the measurement by a factor of $\sqrt{N}$, where $N$ would correspond to the number of stacked galaxies, and provides a representative overview of the general properties of the galaxies in the sample, at the cost of losing the ability to characterize peculiarities from individual objects.
\\
In a recent paper, \citet{Guo2023b} stack the MUSE data cube segments of a sample of 172 galaxies from the MUSE Hubble Ultra Deep Field surveys at $z\approx1$, after separating them as edge-on or face-on, and aligning the edge-on subsample along the direction of their photometric major axis, as determined from HST imaging. 

They compute \ion{Mg}{II} pseudo-narrowband images for the edge-on and face-on stacked cubes and detect extended \ion{Mg}{II} emission over several kpc for both subsamples. In the case of edge-on galaxies, they find a clear enhancement of \ion{Mg}{II} emission along the minor axis, which suggests a bipolar geometry, extending out to a radius of over 10 kpc, with little absorption in the central region. For face-on galaxies, on the other hand, they detect weaker extended emission, but with stronger central absorption, surrounded by a ring-like emission pattern, whose physical origin remains unclear but could be attributed to an isotropic \ion{Mg}{II} halo, inflowing or re-accreted gas, or to a face-on outflow cone extending to even larger radii.

A strong central absorption surrounded by a ring-like pattern qualitatively corresponds to what we see in Fig.~\ref{fig:cont_subs_FoV}, for both the data and best-fitting model cubes. In the case of the model, this configuration results purely as a consequence of the outflow cone being seen nearly face-on (see also Fig.~\ref{fig:bicone_best_fit} for a schematic representation of the best-fitting model geometry). However, as established in previous sections, our model underestimates the total amount of emission in the outer apertures (beyond $\sim9$ kpc). Hence, although we cannot rule out a contribution of other physical processes to the ring-like emission pattern seen in the data cube, a face-on biconical outflow is perfectly consistent and can, to first order, explain this feature.

Regarding the outflow properties, \citet{Guo2023b} find their data to be consistent with an average opening angle of $\sim70\pm11^{\circ}$, in the case of the edge-on subsample, which is relatively close to our measurement of $\sim59\pm4^{\circ}$, especially considering the radically different approaches to measure these values, which in the case of \citet{Guo2023b} was measured from the pseudo-narrowband images. They estimate an outflow velocity of $180$ km s$^{-1}$. This velocity corresponds to the velocity of the outflow at a radius of $\sim10$ kpc, which in the case of our best-fitting model would be of $120\pm30$ km s$^{-1}$, in reasonable agreement, within two standard deviations, which is remarkable considering the completely different nature of both approaches.

Finally, \citet{Guo2023b} infer an outflow rate of $\dot M$ = $36.3\substack{+16.4 \\ -12.4}$ M$_{\odot}$ yr$^{-1}$, which is higher than our value reported in Sec~\ref{sec:outflow_mass_rate}. However, as discussed in Sect.~\ref{sec:outflow_mass_rate}, if the difference in the assumed dust depletion factor is accounted for, the mass outflow rate of \#884 is actually higher than their measurement. 

Nevertheless, we remind the reader that while \citet{Guo2023b} infer the average outflow properties of an ensemble of galaxies, many of which exhibit very low surface brightness on their extended \ion{Mg}{II} (some even show a S/N < 0 along the semiminor axis), we measure here the outflow properties of a single galaxy chosen by its remarkably extended \ion{Mg}{II} emission. Hence, it is likely not representative of the average properties of galaxies. In this sense, it is hardly surprising that our measured outflow mass rate is higher than their estimation (once the different assumptions are accounted for), since our galaxy could perfectly be on the high part of the outflow mass rate distribution. However, a systematic study on an individual-galaxy basis is needed in order to assess this statement quantitatively.

\section{Summary and conclusions}
\label{sec:summary}

We have presented the discovery of a star-forming galaxy surrounded by a very extended \ion{Mg}{II} emission that can be traced out to 30--40~kpc ($\ga$20 half-light radii) from the center of a galaxy at $z=0.737$, with a stellar mass $\log (M_\star$/M$_\odot) = 10.3 \pm 0.3$ and $\mathrm{SFR} \simeq 10\pm7\:\mathrm{M}_\odot\:\mbox{yr}^{-1}$. The system constitutes a great opportunity to study the circumgalactic medium in emission. We performed a modeling of the \ion{Mg}{II} halo to shed light on the physics at play in the CGM, and our main findings are the following:

\begin{enumerate}
\setlength{\itemsep}{0.5ex}

\item The galaxy spectrum itself is dominated by \ion{Mg}{II} in absorption, albeit with weak P-Cygni emission humps. Toward larger impact parameters, the circumgalactic emission becomes visible and eventually dominant (Fig.~\ref{fig:aperspec1}) at surface brightnesses around and below $10^{-18}$ erg s$^{-1}$ cm$^{-2}$ arcsec$^{-2}$ (Fig.~\ref{fig:mg2nb1}), up to $\sim30-40$ kpc from the central galaxy.

\item We used a simple outflow modeling scheme, where we consider the outflow to be composed of thin spherical shells and whose velocity increases with radii, to fit the inner 26 kpc of the \ion{Mg}{II} halo emission of \#884 in a spatially resolved manner. We used an MCMC approach to find the best-fitting parameters, and we produced a model cube that we compared with our data cube.

\item Despite its simplicity, the model is able to reproduce several key features of the \ion{Mg}{II} emission halo present in the data, especially in the inner regions of the CGM. The spectra extracted from ring-like apertures in the model are consistent with those extracted from the data cube, providing an accurate representation of how the P-Cygni profile changes with impact parameter to the central galaxy. Furthermore, a comparison of the radial gradients of the absorption and emission components of the P-Cygni profile shows that the modeled halo also exhibits a spatial distribution consistent with that present in the data.

\item We were able to constrain the velocity profile with which the circumgalactic gas expands from the central source, described by the parameter $\gamma$, the slope of the power law dependence of velocity on radius. We find that our best model is given by $\gamma = 0.7\pm 0.1$, implying that the gas velocity increases steeply toward larger distances from the central galaxy. This also demonstrates that using the Sobolev approximation to solve the radiative transfer problem is a self-consistent choice.

\item The model also provides constraints on the launching velocity of the wind ($v_{0}$), the maximum velocity (or radius) reached by the outflow, and the central optical depth (and thus gas density).

\item In terms of geometry, we find that the outflow is seen nearly face-on (i.e., the bicone points toward the observer, a rotation angle of $80\pm10$), with an opening angle of $59\pm4$. This geometry agrees well with what we would expect from a bipolar outflow given that the central galaxy is also seen nearly face-on.

\item We find that the terminal velocity of the outflow is considerably higher than the escape velocity of the halo in which \#884 resides, implying that the outflowing material will likely escape the gravitational potential of the galaxy and enrich the intergalactic medium, rather than falling back onto \#884. We estimate a mass outflow rate of $12\pm7\,\mathrm{M}_{\odot} \,\mathrm{yr}^{-1}$, which implies a mass loading factor of approximately one. However, we emphasize that these numbers are subject to significant systematic uncertainties that come from the assumptions made to estimate them.

\item There are a number of differences between the data and the model. These discrepancies arise due to the difficulty for our outflow model to faithfully reproduce higher-order features of the \ion{Mg}{II} emission, such as its complex kinematic structure in the outer regions. These differences point out that mechanisms other than outflows, such as inflows, satellites, or turbulence, could become relevant (or dominant) in shaping the CGM of galaxies, particularly toward large impact parameters.


\end{enumerate}

We have shown that a relatively simple outflow modeling scheme is able to reproduce many key features of the \ion{Mg}{II} halo around \#884 and provide constraints on relevant physical mechanisms, such as the launching velocity of the galactic wind and its velocity gradient, which ultimately can be critical to understanding the interplay between galaxies and their CGM. Our model reproduces the data more faithfully in the inner apertures, where the contribution from outflows is expected to be dominant. Some differences between the data and the model toward the outer apertures might indicate an increasingly relevant contribution of other phenomena shaping the CGM. The clear similarity of the circumgalactic \ion{Mg}{II} halo around \#884 to the extended Lyman-$\alpha$ envelopes of high-redshift galaxies encourages the use of these \ion{Mg}{II} haloes at intermediate redshifts as an analog to study and better understand the CGM emission from high-$z$ galaxies. 

It would also be interesting to explore whether our simple modeling approach is also suitable for outflows with orientation angles other than nearly face-on and, in more general terms, for a more diverse set of outflows. Alternatively, it would be of interest to investigate if implementing a more sophisticated modeling approach is worthwhile, such as proper radiative transfer modeling, in order to faithfully reproduce the \ion{Mg}{II} emission of a more representative sample of galaxies.

In this context, available deep VLT/MUSE data give us the opportunity to study galactic outflows in a more systematic fashion. We have already found some other galaxies that present extended \ion{Mg}{II} emission, and in a future paper, we plan to extend our modeling to a larger sample of galaxies and to cover a larger region of the parameter space of galaxy properties, such as stellar mass and redshift. In this way, we seek to provide statistically significant constraints to the CGM physics and the mechanisms that shape it.

\begin{acknowledgements}

IP, LW, and DK acknowledge funding by the European Research Council through ERC-AdG SPECMAP-CGM, GA 101020943.
HS and LW acknowledge funding by the Deutsche Forschungsgemeinschaft, Grant Wi 1369/32-1.
JP and LW acknowledge funding by the Deutsche Forschungsgemeinschaft, Grant Wi 1369/31-1.
HK acknowledges support from the Japan Society for the Promotion of Science (JSPS) Research Fellowships for Young Scientists.

\end{acknowledgements}

\bibliographystyle{aa}
\bibliography{mg2halo}

\begin{appendix}
\section{Aperture-integrated spectrum modeling}
\label{sec:aperture_integrated}
In this section, we compare our results with those obtained using the same methodology described in Sect.~\ref{sec:fitting_method}, but modeling the aperture-integrated spectrum of the data cube instead of doing it in a spatially resolved manner. This is achieved simply by integrating in the spatial dimensions the data and model cubes, before calculating the likelihood, that is, computing the likelihood using only a 1D spectrum.

\begin{figure*}
\includegraphics[width=\textwidth]{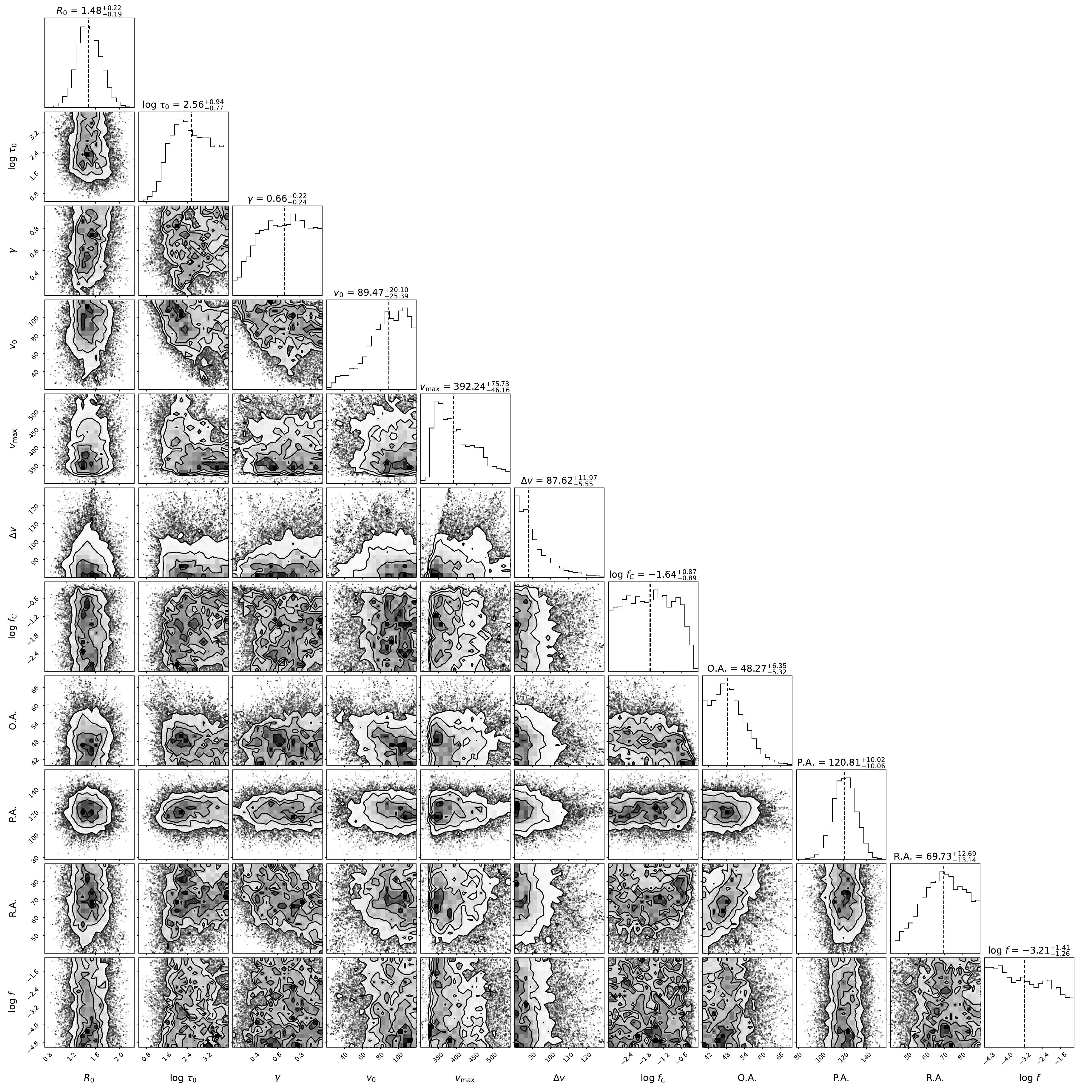}
\caption{Same as Fig.~\ref{fig:posterior_dist}, but for the aperture integrated modeling case (see Appendix~\ref{sec:aperture_integrated}).}
\label{fig:posterior_dist_full_ap}
\end{figure*}

Figure~\ref{fig:posterior_dist_full_ap} shows the posterior distribution of the model parameters obtained when we fit the aperture-integrated spectrum. Most parameters are very poorly constrained, compared to the fiducial spatially resolved modeling. Interestingly, the optical depth $\tau_{0}$ is relatively well constrained (albeit with large uncertainties), with a maximum a posteriori of $400 \pm 600$. This is not unexpected, since while $\tau_{0}$ is the most impactful in the shape of a single aperture-integrated spectrum, the rest of the parameters are constrained by the spatially resolved features of the emission, for instance, how steep is the profile, in the case of $\gamma$. However, we note that in the context of modeling the aperture-integrated spectrum, $\tau_{0}$ does not represent the optical depth at $R_{0}$, but some average quantity across the full FoV. Hence, it is significantly lower than the inferred value from our fiducial analysis.

\begin{figure*}
\includegraphics[width=\textwidth]{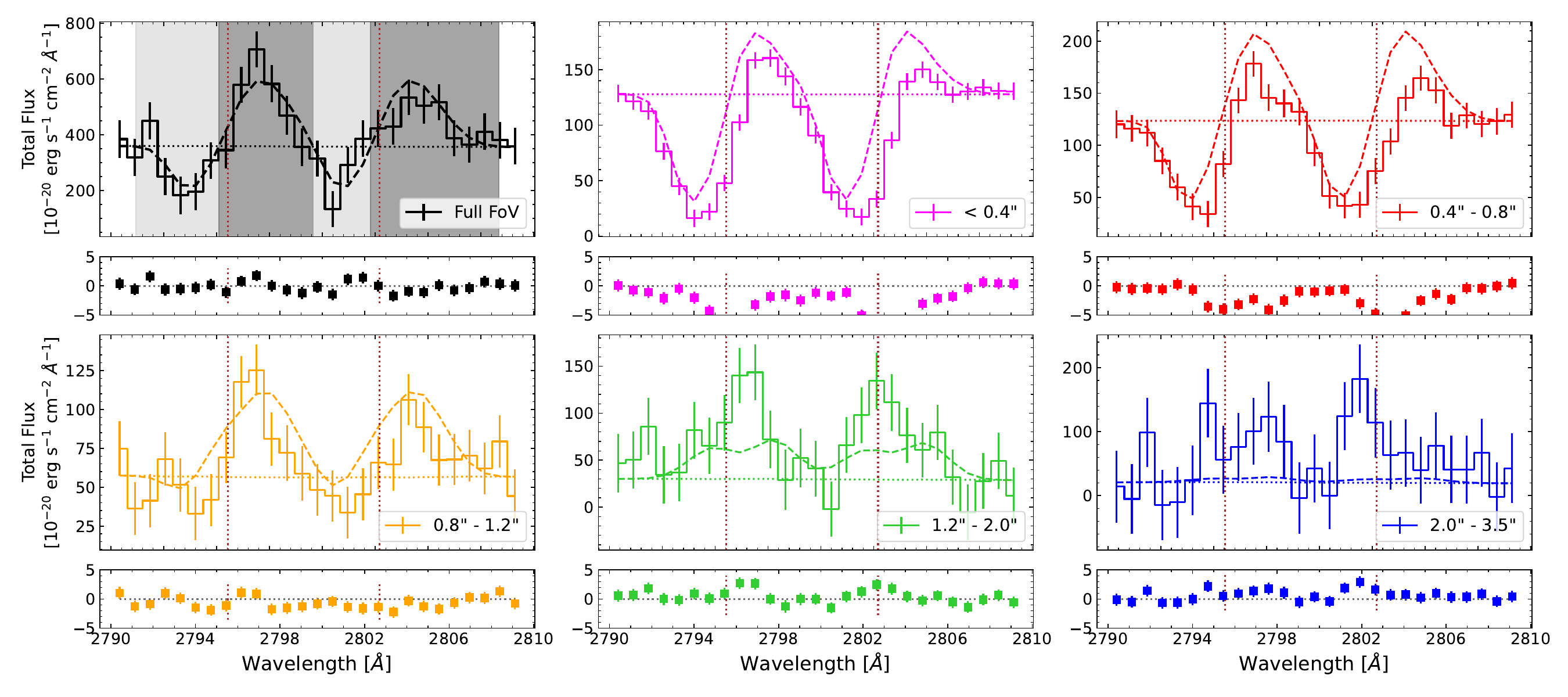}
\caption{Same as Fig.~\ref{fig:annular_spectra}, but for the aperture integrated modeling case (see Appendix~\ref{sec:aperture_integrated}).}
\label{fig:annular_spectra_full_ap}
\end{figure*}

Figure~\ref{fig:annular_spectra_full_ap} shows the comparison between the annular aperture spectra extracted for the model and data cubes (same as Fig.~\ref{fig:annular_spectra}), for the aperture-integrated modeling case. It is clear that the best-fitting model successfully reproduces the integrated \ion{Mg}{II} emission (black). However, a closer examination shows that while the total model emission/absorption matches the data fairly well (and thus, the integrated spectral shape depends primarily on $\tau_{0}$), the spatial distribution of light differs significantly between both cubes. This is an example of possible degeneracies when fitting total integrated light, and highlights the relevance of having a spatially resolved dataset to produce a more accurate representation of the CGM physics.

\FloatBarrier

\section{Central aperture modeling}
\label{sec:inner_aperture_modeling}
\begin{figure}
\includegraphics[width=0.9\columnwidth]{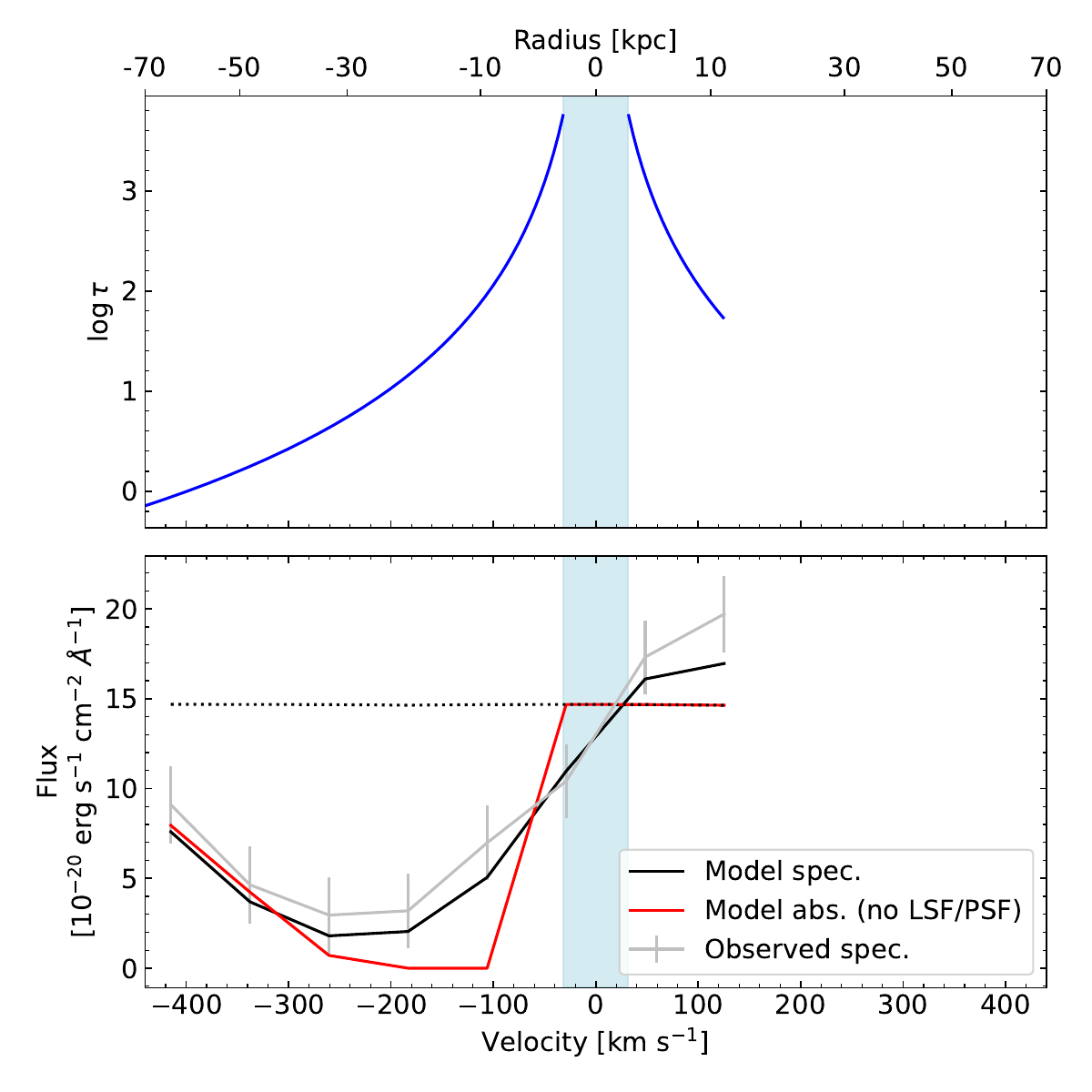}
\caption{Relation between radius, velocity, flux, and optical depth, for the spectrum of the central spaxel of the model and data cubes,  for the best-fitting parameters determined in Sect.~\ref{sec:best_fitting_model}. Top panel shows how the optical depth $\tau$ is higher in the central region of the outflow, and decreases toward larger radii. The light blue area marks the region where radius < $R_{0}$. The bottom panel shows the spectrum that results from the $\tau$ radial gradient shown above. The absorption is strongest closer to zero velocity, where the optical depth is higher, and then its strenght decreases toward more negative velocity, that correspond to larger radii and lower optical depths. Within the light blue band there is no outflow and hence no absorption.   The figure shows the spectrum of only the \ion{Mg}{II} $\lambda$2796 transition, and the red wing of the emission line is truncated to not include wavelength channels where the absorption of the \ion{Mg}{II} $\lambda$2803 transition becomes relevant, making more difficult the interpretation of the figure.}
\label{fig:inner_pixel_modeling}
\end{figure}

As discussed in Sect.~\ref{sec:best_fitting_model}, the blue wing of the \ion{Mg}{II} absorption in the innermost aperture is particularly sensitive to the radial gradient of $\tau$, as this aperture provides a down-the-barrel view of the outflow, and changes in $\tau$ along the line of sight dominate the resultant line profile, without any additional contribution from the emission component of the P-Cygni profile. Figure~\ref{fig:inner_pixel_modeling} shows explicitly the relation between radius, velocity, flux, and optical depth, for the spectrum of the central spaxel of the model and data cubes, for the best-fitting parameters determined in Sect.~\ref{sec:best_fitting_model}.

\begin{figure*}
\includegraphics[width=\textwidth]{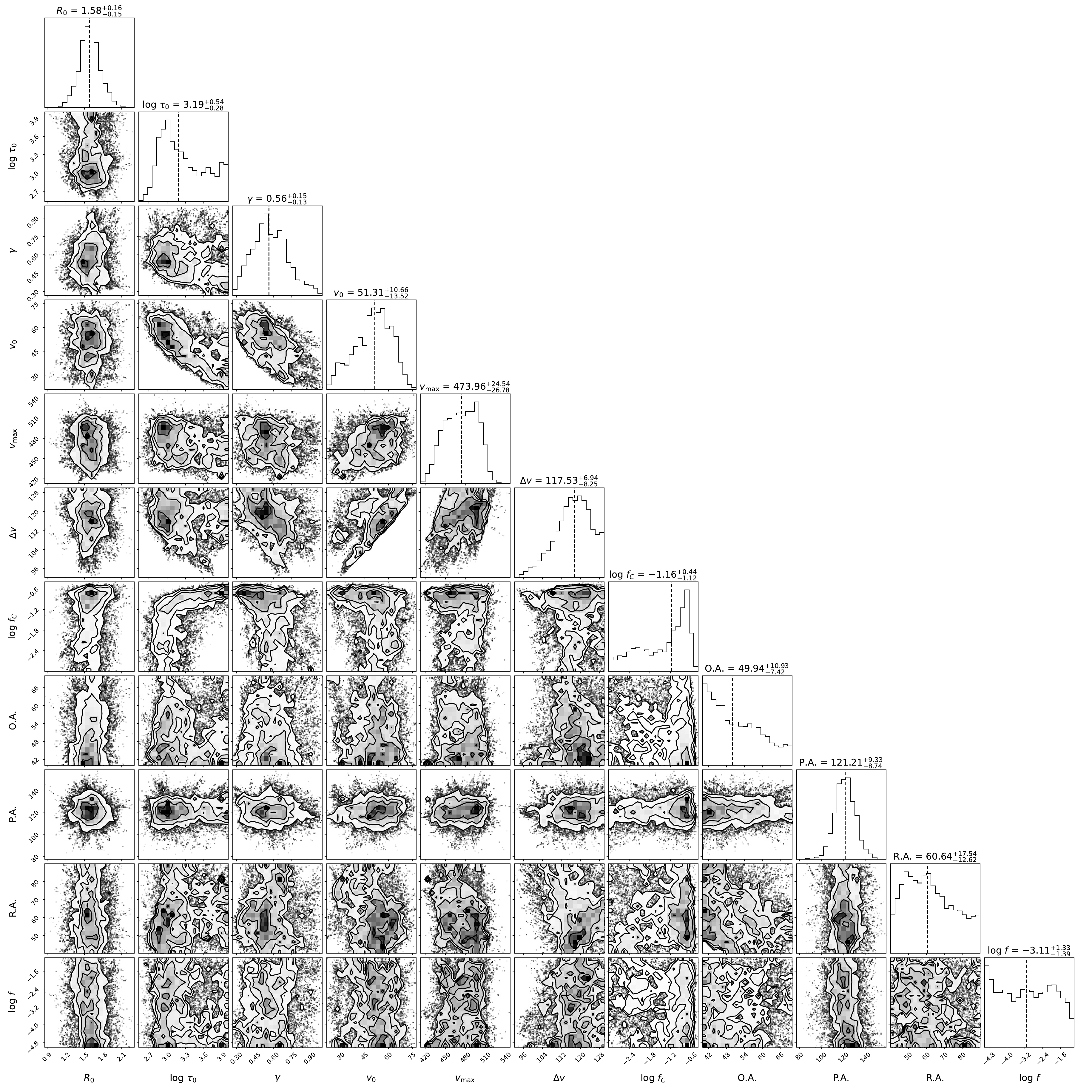}
\caption{Same as Fig.~\ref{fig:posterior_dist}, but for the central aperture modeling case (see Appendix~\ref{sec:inner_aperture_modeling}).}
\label{fig:posterior_dist_central_ap}
\end{figure*}

For comparison purposes, in this section we report the best-fitting parameters obtained when we model only the flux coming from the innermost aperture (magenta circumference in Fig.~\ref{fig:mg2nb1}) of the MUSE data cube. Figure~\ref{fig:posterior_dist_central_ap} shows the posterior distribution of the model parameters obtained in this case. Notably, we are still able to provide constraints in the wind properties when we only use this small portion of the data cube, that are roughly consistent with the numbers reported in our fiducial approach (see Sect.~\ref{sec:best_fitting_model}). This can be explained because although we are observing only a small region, given that the outflow is seen nearly face-on, the central sight line alone still provides information along the full outflow.

The optical depth inferred in this scenario is somewhat lower than that measured in a larger FoV, and we interpret this as a consequence of the excess of emission in the outer apertures driving $\tau_0$ toward higher values. Due to covariance effects, a smaller $\tau_{0}$ also leads to a slightly smaller $\gamma$ and larger $v_{0}$, compared to our fiducial values. On the other hand, unsurprisingly, the geometry of the outflow is very poorly constrained from modeling only the central sight line.

We have also attempted to model the full cube while keeping the value of $\tau_{0}$ fixed to the maximum posterior value obtained in this test, but this does not provide a clear improvement in the modeling of the data cube, nor in the constrain of the model parameters. Moreover, the derived parameters in this case are generally consistent with those reported in Sect.~\ref{sec:best_fitting_model}. Thus, we opt to keep $\tau_{0}$ as a free parameter in our fiducial modeling. Nevertheless, we conclude that, at least for a face-on outflow, modeling exclusively the central sight line could provide some meaningful constraints in the wind properties. 

This contrasts with what we report in Appendix~\ref{sec:aperture_integrated}, that a full aperture-integrated modeling does not provide significant constraint to the outflow parameters, and might look like adding information has a negative impact on the inference of physical parameters. However, the key difference is that in the full aperture-integrated modeling, the model fitting completely loses track of where the photons come from because photons at a given velocity (and thus distance from the central source) could come either from a larger impact parameter that is closer in the line of sight direction to the central source or from a smaller impact parameter that is farther away in the line of sight direction. In other words, distances in the plane of sky and line of sight direction cannot be disentangled. On the other hand, when modeling only the central region, different velocities can only correspond to different locations along the line of sight. Thus, paradoxically, we can track the origin of the photons better and hence infer the outflow properties by modeling the central sight line, provided that the outflow is (nearly) face-on. Otherwise, the central sight line would only probe a limited portion of the outflow in the velocity space.

\end{appendix}

\end{document}